\documentclass[12pt, prd, showpacs]{revtex4-1}
\usepackage{amssymb,amsmath,mathtools,xcolor,graphicx,xspace,colortbl,ragged2e,rotating}
\usepackage{textcomp}
\usepackage{amsfonts}
\usepackage{comment}
\usepackage{amsmath}
\usepackage{amssymb,amsmath,mathtools,xcolor,graphicx,xspace,colortbl,ragged2e,rotating}
\usepackage{boxedminipage}
\usepackage{graphics}
\usepackage{ragged2e}
\usepackage{tabulary}
\usepackage{wrapfig}
\usepackage{xcolor}
\usepackage{multirow}
\usepackage{booktabs}
\usepackage{tabularx,ragged2e,booktabs}
\usepackage{multirow}
\usepackage[paper=letterpaper, twoside=false,textwidth=7.1in, textheight=8.7in,left=0.7in, top=0.99in,headheight=0.17in, headsep=0.17in,]{geometry}

\newcolumntype{C}{>{\Centering\arraybackslash}X}

\begin{document}

\title{Ba\~{n}ados-Silk-West effect with finite forces near different types
of horizons: general classification of scenarios}
\author{H. V. Ovcharenko,}
\email{gregor\_ovcharenko@outlook.com}
\author{O. B. Zaslavskii}
\email{zaslav@ukr.net}

\begin{abstract}
If two particles move towards a black hole and collide in the vicinity of
the horizon, under certain conditions their energy $E_{c.m.}$ in the center
of mass frame can grow unbounded. This is the Ba\~{n}ados-Silk-West (BSW)
effect. Usually, this effect is considered for extremal horizons and
geodesic (or electrogedesic) trajectories. We study this effect in a more
general context, when both geometric and dynamic factors are taken into
account. We consider generic axially symmetric rotating black holes. The
near-horizon behavior of metric coefficients is determined by three numbers $%
p,~q,$ $k$ that appear in the Taylor expansions for different types of a
horizon$.$ This includes nonextremal, extremal and ultraextremal horizons.
We also give general classification of possible trajectories that include
so-called usual, subcritical, critical and ultracritical ones depending on
the near-horizon behavior of the radial component of the four-velocity. We
assume that particles move not freely but under the action of some
unspecified force. We find when the finiteness of a force and the BSW effect
are compatible with each other. The BSW effect implies that one of two
particles has fine-tuned parameters. We show that such a particle always
requires an infinite proper time for reaching the horizon. Otherwise, either
a force becomes infinite or a horizon fails to be regular. This realizes the
so-called principle of kinematic censorship that forbids literally infinite $%
E_{c.m.}$ in any act of collision. The obtained general results are
illustrated for the Kerr-Newman-(anti-)de Sitter metric used as an example.
The description of diversity of trajectories suggested in our work can be of
use also in other contexts, beyond the BSW effect. In particular, we find
the relation between a force and the type of a trajectory. \newline
\end{abstract}

\maketitle

\affiliation{Department of Physics, V.N.Karazin Kharkov National University, 61022
Kharkov, Ukraine}

\affiliation{Charles University, Faculty of Mathematics and Physics, Institute of
Theoretical Physics, V Holesovickach 2, 18000 Prague 8, Czechia.}

\affiliation{Department of Physics and Technology, Kharkov V.N. Karazin National
University, 4 Svoboda Square, Kharkov 61022, Ukraine}

\section{Introduction}

The Ba\~{n}ados, Silk and West effect (BSW, after the names of its authors) 
\cite{ban} is one of the most interesting theoretical results in black hole
physics during the last decade. It also revived interest to previous
versions of high energy collisions near black holes \cite{pir1} - \cite{pir3}%
. Let two particles collide in the vicinity of a rotating black hole. Then,
under certain conditions, an indefinitely growth of the energy in the center
of mass frame $E_{c.m.}$ becomes possible. This effect was found for (i)
extremal horizons and (ii) free particles. Some objections against the BSW
effect \cite{berti}, \cite{ted} were connected with failure of the factors
(i), (ii) or both. However, it was shown later, that under some change of
conditions, the BSW effect survives even for nonextremal black holes \cite%
{gp}. Moreover, it was shown earlier that the BSW effect arises due to the
presence of the horizon as such, no matter how its explicit metric looks
like \cite{prd}. In the present work me make the next step and consider
generic horizons including nonextremal, extremal and utraextremal ones (more
explicit definitions will be done in the text below). For spherically
symmetric space-times there exists their direct classification that enables
us to distinguish between true regular horizons, light-like singularities
and so-called naked and truly naked horizons (see \cite{prd08} and
references therein). For generic axially symmetric rotating black holes
classification is much more complicated. The conditions that single out
standard regular horizons (which we restrict ourselves by) were described in 
\cite{ov-zas}.

Also, the presence of a force can be, in principle, compatible with the BSW
effect. For a particular case of extremal horizons this was shown in \cite%
{tz13}. Strong arguments in favour of this effect for nonextremal horizons
were suggested in \cite{tz14}. Moreover, sometimes it leads to another
version of this effect which is absent without a force \cite{ac}.

Instead of solving the equations of motion (that, as a rule, is practically
impossible) we choose the near-horizon behavior of trajectories and find for
each type of a horizon, when (i) the acceleration due to a force remains
finite near the horizon and, at the same time (ii) the BSW effect is
allowed. For particles moving in the equatorial plane toward a black hole
and experiencing finite forces, we build a general theory of the BSW effect.
In doing so, we take into account factors connected with geometry (type of a
horizon), kinematics (classification of trajectories) and dynamics (allowed
behavior of a force).

One important aspect deserves separate attention. Although $E_{c.m.}$ can be
made as large as one likes, if the BSW effect is present, its value must
remain finite in each act of collision, so an infinite energy is forbidden.
This statement is formulated as a separate principle of kinematic censorship 
\cite{cens}. As far as the BSW effect with free moving particles is
concerned, it implies that one of two colliding particles has fine-tuned
parameters, then the the proper time required to reach the extremal \
horizon is infinite \cite{ted}. Thus collision occurs closely to the horizon
but not exactly on it, so that $E_{c.m.}$ remains finite, although
arbitrarily large. We show how this principle manifests itself for more
general types of horizons and the presence of a nonzero force.

The paper is organized as follows. In Sec. \ref{gen-set} we write the
general form of the metric under discussion and equations of particle motion
under the action of a nonzero force. In Sec. \ref{four-vel} we suggest
classification of trajectories depending on the near-horizon behavior of the
radial component of the four-velocity. In this way, we introduce notions of
usual, subcritical, critical and ultracritical particles. In Sec. \ref{beh}
we establish main features of different types of trajectories in the
vicinity of the black hole horizon. In Sec. \ref{energ} we give the basic
formulas for the gamma factor of relative motion of two particles relevant
in the context of the BSW effect. We enumerate different possible
combinations of types of both particles that produce the BSW effect. In Sec. %
\ref{expr} we list general expressions for the components of acceleration
for equatorial particle motion. In Sec. \ref{fin-tun} we establish the
relations between the type of trajectory, acceleration and characteristics
of near-horizon metric. We derive the conditions when the corresponding
force is finite for fine-tuned particles. In Sec. \ref{us}, we derive, for
completeness, similar conditions for usual particles, although this is
irrelevant for the conditions of the BSW effect. In Sec. \ref{results_typ}
we collect our results about conditions when a force remains finite near the
horizon for fine-tuned particles and different kinds of horizons. In Sec. %
\ref{proper} we prove the validity of the principle of kinematic censorship
for the system under discussion. Then, in Sec. \ref{check-res} we check the
validity of our results using the Kerr-Newman-(anti-)de Sitter metric as an
exactly solvable example. In Sec. \ref{sum} we give the summary of the
results obtained in this work.

\section{Metric and equations of motion}

\label{gen-set}

We investigate the motion of particles in the background of a rotating black
hole described in generalized Boyer-Lindquist coordinates by the metric 
\begin{equation}
ds^{2}=-N^{2}dt^{2}+g_{\varphi \varphi }(dt-\omega d\varphi )^{2}+\dfrac{%
dr^{2}}{A}+g_{\theta \theta }d\theta ^{2}.  \label{metr}
\end{equation}

All metric coefficients do not depend on $t$ and $\varphi $. Positions of
horizons are defined by the conditions $A(r_{h})=N(r_{h})=0$, where $r_{h}$
is the horizon radius.

The BSW phenomenon supposes that the energy $E_{c.m.}$ in center of mass
frame of two colliding particles infinitely grows as the point of collision
approaches the black hole horizon. For the extremal horizon, the parameters
of one of particles (so-called critical) should be fine-tuned, the other
particle being not fine-tuned (usual) \cite{ban}, \cite{prd}. Meanwhile, for
more general types of the horizon the situation can be more involved, as we
will see it below. Choosing a general type of a trajectory, we relate it to
the properties of the horizon and will see how the near-horizon behavior of
acceleration looks like.

Let $\xi ^{\mu }$ and $\eta ^{\mu \text{ }}$be the Killing vectors
responsible for time translation and rotation around the axis, respectively.
Then, one can introduce the energy $E=-mu_{\mu }\xi ^{\mu }=-mu_{t}$ and
angular momentum $L=mu_{\mu }\eta ^{\mu }=mu_{\varphi }$, where $u^{\mu }$
is the four-velocity, $m$ being a particle's mass. It follows that along the
particle trajectory the derivative with respect to the proper time $\tau $
gives us%
\begin{equation}
\frac{d\varepsilon }{d\tau }=-a_{\mu }\xi ^{\mu }\text{,}
\end{equation}%
\begin{equation}
\frac{d\mathcal{L}}{d\tau }=a_{\mu }\eta ^{\mu }\,,
\end{equation}%
where the four-acceleration%
\begin{equation}
a_{\mu }=u_{\mu ;\nu }u^{\nu }\text{,}  \label{a}
\end{equation}%
semicolon denotes covariant derivative, $\varepsilon =\frac{E}{m},$ $%
\mathcal{L=}\frac{L}{m}$. For a free particle, $a_{\mu }=0$ and the energy
and angular momentum are conserved.

Hereafter, we assume that the metric possesses a symmetry with respect to
the equatorial plane $\theta =\frac{\pi }{2}$ and restrict ourselves by
particle motion in this plane. Then, using the definitions of $\varepsilon $
and $\mathcal{L}$ and the normalization condition $g_{\mu \nu }u^{\mu
}u^{\nu }=-1$, one can find that%
\begin{equation}
u^{t}=\frac{\mathcal{X}}{N^{2}}\text{,}  \label{t}
\end{equation}%
\begin{equation}
u^{\varphi }=\frac{\mathcal{L}}{g_{\varphi \varphi }}+\frac{\omega \mathcal{X%
}}{N^{2}}\text{,}  \label{uphi}
\end{equation}%
\begin{equation}
u^{r}=\sigma \frac{\sqrt{A}}{N}P\text{,}  \label{ur}
\end{equation}%
\begin{equation}
P=\sqrt{\mathcal{X}^{2}-N^{2}(1+\frac{\mathcal{L}^{2}}{g_{\varphi \varphi }})%
}\text{,}  \label{P}
\end{equation}%
\begin{equation}
\mathcal{X}=\varepsilon -\omega \mathcal{L}\text{,}  \label{hi}
\end{equation}%
where $\sigma =\pm 1$ depending on the direction of motion. As in our work
we restrict ourselves by motion within the equatorial plane, the component $%
u^{\theta }=0$.

If $a_{\mu }\neq 0$, $\varepsilon $ and $\mathcal{L}$ are not conserved but
eqs. (\ref{t}) - (\ref{ur}) hold true anyway.

It follows from eqs. (\ref{t}) - (\ref{ur}) that in coordinates $(t,\varphi
,r,\theta )$ 
\begin{equation}
u^{\mu }=\Big(\dfrac{\mathcal{X}}{N^{2}},\dfrac{\mathcal{L}}{g_{\varphi }}+%
\dfrac{\omega \mathcal{X}}{N^{2}},\sigma \dfrac{\sqrt{A}}{N}P,0\Big).
\label{u}
\end{equation}

In what follows, we will use, along with the coordinate components of
vectors, also their tetrad components. It is convenient to introduce the
tetrad attached to the so-called zero angular momentum observers (ZAMO)
according to \cite{72}. This tetrad reads in our coordinates 
\begin{eqnarray}
e_{\mu }^{(0)} &=&N(1,0,0,0),~~~e_{\mu }^{(1)}=\sqrt{g_{\varphi \varphi }}%
(-\omega ,1,0,0),~~~  \label{zamo} \\
e_{\mu }^{(2)} &=&\dfrac{1}{\sqrt{A}}(0,0,1,0),~~~e_{\mu }^{(3)}=\sqrt{%
g_{\theta \theta }}(0,0,0,1).
\end{eqnarray}%
We will use a letter "O" (orbital) to call them OZAMO to stress that for
such an observer $r=const$. Trajectories of this kind of observers are, in
general, not geodesics in contrast to FZAMO (free-falling observers with a
zero angular momentum).

\section{Four-velocities and classification of trajectories}

\label{four-vel}

Hereafter, we will use the following classification of particles
(trajectories) depending on the near-horizon behavior of $u^{r}$. Let $%
N\rightarrow 0$, $A\rightarrow 0$. Then, we call a particle usual if $%
\left\vert u^{r}\right\vert \approx \dfrac{\sqrt{A}}{N}$ , subcritical if $%
\left\vert u^{r}\right\vert $ changes slower than $\sqrt{A}$ but faster than 
$\dfrac{\sqrt{A}}{N}$, critical if $\left\vert u^{r}\right\vert \approx 
\sqrt{A}$, ultracritical if $\left\vert u^{r}\right\vert $ changes faster
than $\sqrt{A}$.

The standard approach to investigation of particle trajectories consists in
study, how the presence of external forces affects particle dynamics.
Instead of solving this problem, we proceed in the opposite direction: we
set the near-horizon trajectory, calculate acceleration and elucidate when
it is finite. Further, we select the trajectories that give simultaneously
(i) finite acceleration and (ii) divergent $E_{c.m.}$

Afterwards, we are left with the angular component $u^{\varphi }$ and the
time one $u^{t}$. It is seen from (\ref{uphi}) that in the rotational
background (\ref{metr}) the angular component of velocity consists of two
terms. The first one is due to the angular momentum and is related to
rotation itself, the second term appears due to frame-dragging. The second
term is divergent near the horizon, so it is more natural to define the
angular component of the four-velocity in the OZAMO frame (\ref{zamo}): 
\begin{equation}
u_{o}^{\varphi }=e_{\mu }^{(1)}u^{\mu }=\sqrt{g_{\varphi \varphi }}%
(u^{\varphi }-\omega u^{t})=\frac{\mathcal{L}}{\sqrt{g_{\varphi \varphi }}},
\label{uL}
\end{equation}%
which is free from this divergence.

The time component $u^{t}$ is given by eq. (\ref{t}) and can be also
rewritten in another quite convenient form in terms of $u^{r}$ and $%
u_{o}^{\varphi }$. It follows from the normalization condition and (\ref{ur}%
) that%
\begin{equation}
u^{t}=\dfrac{1}{N}\sqrt{1+\dfrac{(u^{r})^{2}}{A}+(u_{o}^{\varphi })^{2}}.
\label{u_t}
\end{equation}

According to (\ref{t}), $\mathcal{X}=u^{t}N^{2}$. Combining this with (\ref%
{u_t}) and taking into account that $u_{o}^{\varphi }=O(1)$, we see that $%
\mathcal{X}_{H}\neq 0$ for usual particles and $\mathcal{X}_{H}=0$ for other
types (subcritical, critical and ultracritical). Hereafter, subscript "H"
denotes the quantities calculated on the horizon.

It also follows from our classification that near the horizon 
\begin{equation}
u^{t}\approx \frac{1}{N^{2}}\text{ for usual particles, }u^{t}\approx \frac{%
u^{r}}{N\sqrt{A}}\text{ for subcritical ones, }u^{t}\approx \frac{1}{N}\text{
for critical and ultracritical}.  \label{utr}
\end{equation}

Traditionally, the classification of the trajectories is based on the
near-horizon behavior of $\mathcal{X}$, while properties of $u^{r}$ are
derived from this as consequences. Such an approach is convenient when
dealing with usual and critical particles. However, as we are going to
analyze more subtle details of trajectories and include into consideration
subcritical and ultracritical ones, the reverse method (from properties of $%
u^{r}$ to those of $\mathcal{X}$) is more convenient, as we will see it
below. In principle, both approaches are equivalent to each other.

Using our classification and eqs. (\ref{ur}), (\ref{P}), we can derive
important consequences for the relation between $u^{r}$ and $\mathcal{X} $
near the horizon. Namely, for usual and subcritical particles, 
\begin{equation}
P\approx \mathcal{X} -\frac{N^{2}}{2\mathcal{X} }(1+\frac{\mathcal{L}^{2}}{%
g_{\varphi \varphi }})_{H}\text{, }\left\vert u^{r}\right\vert \approx \frac{%
\sqrt{A}}{N}\mathcal{X},  \label{puu}
\end{equation}%
for critical ones,%
\begin{equation}
\mathcal{X} \approx X _{1}N\text{, }P\approx P_{N}N\text{, }\left\vert
u^{r}\right\vert \approx P_{1}\sqrt{A}=\frac{P_{1}}{X_{1}}\frac{\sqrt{A}}{N}%
\mathcal{X} ,  \label{puc}
\end{equation}%
where $X_{1}$ and $P_{1}=\sqrt{X_{1}^{2}-(1+\frac{\mathcal{L}^{2}}{%
g_{\varphi \varphi }})_{H}}$ are constants$.$

For ultracritical particles, 
\begin{equation}
\mathcal{X} \approx (1+\frac{\mathcal{L}^{2}}{g_{\varphi \varphi }})_{H}N%
\text{, }P\approx P_{\delta }N^{1+\delta }\text{, }\delta >0\text{, }
\label{pultra}
\end{equation}%
where $P_{\delta }$ is some another constant,%
\begin{equation}
\left\vert u^{r}\right\vert \approx P_{\delta }\sqrt{A}N^{\delta }\text{. }
\end{equation}

Thus we see that for all particles, except from ultracritical ones, $%
\left\vert u^{r}\right\vert $ has the order $\frac{\sqrt{A}}{N}\mathcal{X} $%
. For ultracritical particles, $\left\vert u^{r}\right\vert \ll \frac{\sqrt{A%
}}{N}\mathcal{X} $.

\section{Behavior of velocity near horizon}

\label{beh}

As the BSW effect happens near the horizon, we will focus on the behavior of
accelerations and velocities in its vicinity. The situation depends strongly
on the type of a horizon. The classification of the horizons is based on a
character of the behavior of geometrical entities in a free-falling frame.
Explicitly, it reveals itself in the type of the near-horizon expansion of
the metric coefficients. Let us write them in a general form ($v=r-r_{h}$): 
\begin{equation}
N^{2}=\kappa _{p}v^{p}+o(v^{p}),~~~A=A_{q}v^{q}+o(v^{q})\text{,}
\end{equation}%
\begin{equation}
\omega =\omega _{H}+\hat{\omega}_{k}v^{k}+...+\hat{\omega}%
_{l-1}v^{l-1}+\omega _{l}(\theta )v^{l}+o(v^{l}),  \label{omk}
\end{equation}%
\begin{equation}
g_{a}=g_{aH}+g_{a1}v+o(v).  \label{ga}
\end{equation}%
Here $a=\theta ,\varphi $, hat means that corresponding quantity does not
depend on $\theta $. It is assumed that $p$, $q$, $k$ are some positive
numbers. By definition, if $p=q=1$, the horizon is nonextremal. If $p\geq 2$
and $q=2$, it is extremal. For $q>2$, it is called ultraextremal. For
nonextremal horizons the surface gravity is not equal to zero, for extremal
and ultraextremal ones it is zero. For more details, see \cite{ov-zas}.

We analyze behavior of accelerations for any type of horizon, so $q,$ $p$
and $k$ are arbitrary. According to the results, obtained in \cite{ov-zas},
the regularity of a horizon requires that 
\begin{equation}
k\geq \Big[\dfrac{p-q+3}{2}\Big],~l\geq p,  \label{k}
\end{equation}%
where $[x]$ means integer part of $x$. In what follows, we tacitly assume
that these and other conditions of regularity \cite{ov-zas} are fulfilled.

These expansions allow us to obtain behavior of $\mathcal{X}$. To this end,
we consider a general behavior of the radial velocity near the horizon in
the form%
\begin{equation}
u^{r}=(u^{r})_{c}v^{c}+o(v^{c}).  \label{c}
\end{equation}
Near the horizon, 
\begin{equation}
\mathcal{X}\approx \mathcal{X}_{s}\nu ^{s}\text{,}  \label{s}
\end{equation}%
where $s=0$ for usual particles and $s>0$ in other cases.

It follows from (\ref{utr}) and (\ref{s}) that for subcritical particles 
\begin{equation}
\mathcal{X}\approx \dfrac{N}{\sqrt{A}}\left\vert u^{r}\right\vert
\rightarrow s=\dfrac{p-q}{2}+c.  \label{X}
\end{equation}

For critical and ultracritical particles $s=p/2$. However for ultracritical
particle exists another restriction. To see it, let us consider the radial
component of the four-velocity. For the ultracritical particle we require
that 
\begin{equation}
(u^{r})^{2}=\dfrac{A}{N^{2}}\Big(\mathcal{X}^{2}-N^{2}\Big(1+\dfrac{\mathcal{%
L}^{2}}{g_{\varphi \varphi }}\Big)\Big)\approx v^{2c},
\end{equation}%
where $c>\dfrac{q}{2}.$

Such a behavior of $u^{r}$ can be obtained only if we impose additional
restrictions on $\mathcal{X}^{2}$. It has to be equal to the second
expression inside the radical (\ref{P}) up to the corrections of a higher
order:%
\begin{equation}
\mathcal{X}^{2}=N^{2}\Big(1+\dfrac{\mathcal{L}^{2}}{g_{\varphi \varphi }}%
\Big)~\text{\textrm{plus}}\mathrm{~}v^{2c+p-q}~\mathrm{terms.}
\label{xi_cond}
\end{equation}

For the angular component of the four-velocity we can write another
near-horizon expansion: 
\begin{equation}
u_{O}^{\varphi }=(u_{O}^{\varphi })_{H}+(u_{O}^{\varphi })_{b}v^{b}+o(v^{b}),
\label{b}
\end{equation}%
where $b>0$.

Then, it follows from (\ref{uL}) that 
\begin{equation}
\mathcal{L}\approx \sqrt{g_{\varphi \varphi }}(u_{O}^{\varphi })_{H}+\sqrt{%
g_{\varphi \varphi }}(u_{O}^{\varphi })_{b}\nu ^{b}.  \label{Lb}
\end{equation}%
In a similar way, we can write%
\begin{equation}
u^{t}\approx v^{-\beta }\text{,}  \label{beta}
\end{equation}%
where for usual particles $\beta =p$, for subcritical ones $\beta =\frac{p+q%
}{2}-c$, for critical and ultracritical $b=\frac{p}{2}$.

It is convenient to summarize the above results in a Table \ref{class_of_tr}.

\begin{table}[]
\centering
\begin{tabular}{|l||c|c|c|c|c|}
\hline
~ & Type & $c$ & $\beta $ & $s$ & $\alpha=c-1$ \\ \hline\hline
1 & Usual & $\dfrac{q-p}{2}$ & $p$ & $0$ & $\dfrac{q-p-2}{2}$ \\ \hline
2 & Subcritical & $\dfrac{q-p}{2}<c<\dfrac{q}{2}$ & $\dfrac{p+q}{2}-c$ & $%
\dfrac{p-q}{2}+c$, $0<s<\dfrac{p}{2}$ & $\dfrac{q-p-2}{2}<\alpha<\dfrac{q-2}{%
2}$ \\ \hline
3 & Critical & $\dfrac{q}{2}$ & $\dfrac{p}{2}$ & $\dfrac{p}{2}$ & $\dfrac{q-2%
}{2} $ \\ \hline
4 & Ultracritical & $c>\dfrac{q}{2}$ & $\dfrac{p}{2}$ & $\dfrac{p}{2}$ & $%
\alpha>\dfrac{q-2}{2}$ \\ \hline
\end{tabular}%
\caption{Characteristics of the near-horizon behavior of $u^{r}$, $u^{t}$, $%
\mathcal{X} $ and the proper time $\protect\tau$. Here the proper time
changes as $\protect\tau\approx v^{-\protect\alpha}$. The value $\protect%
\alpha=0$ means that the proper time logarithmically diverges $\protect\tau%
\sim |\ln v|$.}
\label{class_of_tr}
\end{table}

In this context it is also interesting to discuss behavior of a proper time
near the horizon. Using the definition of the radial component of
4-velocity, we have: 
\begin{equation}
\tau =\int \frac{dr}{u^{r}}\approx \int v^{-c}dv.
\end{equation}%
Thus we see that if $c=1$, then $\tau \sim \left\vert \ln v\right\vert $, if 
$c\neq 1,$ $\tau \approx v^{-\alpha }$, where $\alpha =c-1$. If $c=1$, so $%
\alpha =0$, the proper time diverges logarithmically, $\tau \sim |\ln v|$.
This is the case considered in \cite{ban}, \cite{ted}. The case $c=3/2$, $%
\alpha =\frac{1}{2}$ corresponds to so-called critical particles of class II
considered for the Kerr metric in \cite{kd}. Similar solutions for the
extremal Kerr-Newman metric are discussed in \cite{axis}. For equatorial
motion, the proper time for fine-tuned particles in more general background
is considered in \cite{near} \ (but be aware of typos in eq. 91 there).

The proper time is finite if $c<1$. As for all trajectories which we are
considering, $\dfrac{q-p}{2}\leq c$ (Table \ref{class_of_tr}), the proper
time may be finite only if $q<p+2$. Then it becomes possible for $\alpha $
to be negative.

If $q\geq p+2,$ the proper time diverges for all types of trajectories
including the usual ones. It means that the region from infinity to the
horizon is geodesically complete. Such objects are termed "remote horizons"
in \cite{prd08}.

\section{Energy of collision \label{energ}}

As we mentioned above, we are mainly interested in the possibility of the
BSW phenomenon, which is related to infinite growth of energy in the center
of mass frame of two colliding particles. This energy is given by%
\begin{equation}
E_{c.m.}^{2}=-m_{1}m_{2}u_{1}^{\mu }{u_{2}}_{\mu }=m_{1}m_{2}\gamma ,
\end{equation}%
where $\gamma $ is the Lorentz gamma-factor of relative motion. Substituting
expressions for the four-velocity (\ref{u}) we have%
\begin{equation}
\gamma =\dfrac{\mathcal{X}_{1}\mathcal{X}_{2}-P_{1}P_{2}}{N^{2}}-\dfrac{%
\mathcal{L}_{1}\mathcal{L}_{2}}{g_{\varphi \varphi }}.  \label{gamma_1}
\end{equation}%
Hereafter, we assume that both particles move towards the horizon, so $%
\sigma _{1}=\sigma _{2}=-1$.

The second term is always regular, so we are interested in the near-horizon
behavior of the first one. To this end, let us expand the expression for $P$
(\ref{P}): 
\begin{equation}
P^{2}=\mathcal{X}^{2}-N^{2}\Big(\dfrac{\mathcal{L}^{2}}{g_{\varphi }}+1\Big)%
=(\mathcal{X}_{s}v^{s}+\mathcal{X}_{s+s^{\prime }}v^{s+s^{\prime
}}+...)^{2}-\kappa _{p}v^{p}\Big(\dfrac{L_{H}^{2}}{g_{\varphi H}}+1\Big)+...,
\end{equation}%
where $s^{\prime }$ is some positive number. Now let us find how $P$ behaves
near horizon.

For usual and subcritical particles $N\ll \mathcal{X}$, so we can expand the
square root to obtain: 
\begin{equation}
P=\mathcal{X}-\dfrac{N^{2}}{2\mathcal{X}}\Big(\dfrac{\mathcal{L}^{2}}{%
g_{\varphi \varphi }}+1\Big)+...=\mathcal{X}+O(v^{p-s}).  \label{px}
\end{equation}

In cases of critical and ultracritical particles $\mathcal{X}$ and $N$
decrease with the same rate, so we can write 
\begin{equation}
P=P_{p/2}v^{p/2}+...\approx P_{N}N,~~~P_{p/2}=\sqrt{\mathcal{X}%
_{p/2}^{2}-\kappa _{p}\Big(\dfrac{\mathcal{L}_{H}^{2}}{g_{\varphi H}}+1\Big)}%
,\text{ }P_{N}=\frac{P_{p/2}}{\sqrt{\kappa _{p}}}  \label{Pp}
\end{equation}%
that agrees with (\ref{puc}), (\ref{pultra}). As we will further show, for
our purposes it is sufficient to keep the first term in this expansion.

Now let us analyze behavior of the gamma factor. Firstly, let us suppose,
that each of two particles is usual or subcritical. Using (\ref{px}) we see
that (\ref{gamma_1}) becomes%
\begin{equation}
\gamma =\frac{\mathcal{X}_{1}z_{2}+\mathcal{X}_{2}z_{1}}{2}+O(1),
\label{gamma_111}
\end{equation}%
where $z\equiv \dfrac{1+(\frac{\mathcal{L}^{2}}{g_{\varphi \varphi }})_{H}}{%
\mathcal{X}}$. Taking into account (\ref{s}), we see that 
\begin{equation}
\gamma \approx v^{-|s_{1}-s_{2}|}.  \label{s12}
\end{equation}

Note that gamma factor is regular only if $s_{1}=s_{2}$, in other case it
diverges. This result was obtained earlier in \cite{tz13} for the particular
case of extremal horizons with $p=q=2$ (see Sec. II E there). Meanwhile, now
we see that this result is independent on the type of the horizon.

Let now particle 1 be critical or ultracritical while particle 2 be usual.
Then,%
\begin{equation}
\gamma \approx \dfrac{\mathcal{X}_{2}(\mathcal{X}_{p/2}^{(1)}-P_{p/2}^{(1)})%
}{N\sqrt{\kappa _{p}}}=\dfrac{\mathcal{X}_{2}(\mathcal{X}%
_{p/2}^{(1)}-P_{p/2}^{(1)})}{\kappa _{p}v^{p/2}},  \label{cu}
\end{equation}%
where $\mathcal{X}_{2}=O(1)$.

If particle 1 is critical or ultracritical, while particle 2 is subcritical,
in a similar way we obtain (\ref{cu}) with $s>0$ in (\ref{s}), so%
\begin{equation}
\gamma \approx \dfrac{\left( \mathcal{X}_{2}\right) _{s}(\mathcal{X}%
_{p/2}^{(1)}-P_{p/2}^{(1)})}{\kappa _{p}v^{p/2-s}}.
\end{equation}%
This expression is divergent since according to Table \ref{class_of_tr} $s<%
\frac{p}{2}$ for such particles.

If both particles are critical or ultracritical, then $\mathcal{X}%
_{1}\approx P_{1}\sim N$, $\mathcal{X}_{2}\approx P_{2}\sim N$. Thus gamma
factor is regular.

We can generalize these results in a Table \ref{gamma_table}, where we also
introduced quantity $d$ that shows the rate of divergence of the
gamma-factor $\gamma \sim v^{-d}$. From Table \ref{gamma_table} we can
deduce that the BSW phenomenon happens if for a given type of horizon it is
possible to obtain two particles with different rates of decrease of $%
\mathcal{X}$. Firstly let us discuss this possibility for geodesic motion.

\begin{table}[]
\centering
\begin{tabular}{|c||c|c|c|c|}
\hline
& First particle & Second particle & $d$ & $\gamma$ \\ \hline\hline
1 & Usual & Usual & 0 & R \\ \hline
2 & Usual & Subcritical & $s_2$ & D \\ \hline
\multirow{2}{*}{3} & \multirow{2}{*}{Subcritical} & %
\multirow{2}{*}{Subcritical} & \multirow{2}{*}{$|s_1-s_2|$} & D if $s_1\neq
s_2$ \\ 
~ & ~ & ~ & ~ & R if $s_1= s_2$ \\ \hline
\multirow{1}{*}{4} & Usual or Subcritical & Critical or Ultracritical & $%
p/2-s_1$ & D \\ \hline
\multirow{1}{*}{5} & Critical or Ultracritical & Critical or Ultracritical & 
0 & R \\ \hline
\end{tabular}%
\caption{The possibility of BSW phenomenon for different types of particles.
D means that the gamma-factor diverges, R-means that it is regular.}
\label{gamma_table}
\end{table}

In the case of geodesic motion acceleration is zero, thus motion is defined
only by two conserved quantities (energy and angular momentum) and by metric
functions. In this case $\varepsilon $ and $\mathcal{L}$ in (\ref{hi}) are
constants. It follows from (\ref{omk}) that%
\begin{equation}
\mathcal{X}=\mathcal{X}_{H}-\omega _{k}\mathcal{L}v^{k}+o(v^{k}).
\label{hik}
\end{equation}%
If a particle is fine-tuned, $\mathcal{X}_{H}=0$, $\mathcal{L=}\frac{%
\varepsilon }{\omega _{H}}>0$. This gives us that for fine-tuned particles $%
s=k$. This gives realization of the BSW phenomenon if the first particle is
usual, while the second one is fine-tuned. It is worth noting that in this
case the relation $s=k\leq p/2$ has to hold. It comes from reality of the
radial component of the 4-velocity.

In a general case, when forces are present, the expansion for $\varepsilon $
and $\mathcal{L}$ can, in principle, violate the equality $s=k$.

One reservation is in order. In some cases, the BSW process between a usual
and fine-tuned particles fails because of impossibility for a fine-tuned one
to reach the horizon. In particular, this happens for nonextremal horizons
and geodesic particles. Then, the effect can be saved if one of particles is
not fine-tuned exactly \cite{gp}. When the force is present, this is also
compatible with the BSW effect \cite{tz14}. More general situation, with
arbitrary $p,q$ and the presence of a finite force, requires separate
treatment. In this work we put this issue aside and consider the "pure" BSW
effect only, when one of particles is fine-tuned exactly.

\section{General expressions for acceleration}

\label{expr}

As we will consider particle collisions under the presence of forces, for
further analysis we need to have explicit expressions for the components of
acceleration. They are given in the present subsection. It follows from (\ref%
{a}) that 
\begin{equation}
a^{\mu }=u^{\nu }\partial _{\nu }u^{\mu }+\Gamma _{\nu \sigma}^{\mu }u^{\nu
}u^{\sigma},
\end{equation}%
where $\Gamma _{\nu k}^{\mu }$ are Christoffel symbols.

The tetrad components of acceleration: $a_{o}^{(a)}=a^{\mu }e_{\mu }^{(a)}$
can be found from (\ref{zamo}):%
\begin{equation}
a_{o}^{(t)}=Na^{t}~~~~~~a_{o}^{(r)}=\dfrac{a^{r}}{\sqrt{A}}%
~~~a_{o}^{(\varphi )}=\sqrt{g_{\varphi \varphi }}(a^{\varphi }-\omega
a^{t})~~~~a_{o}^{(\theta )}=\sqrt{g_{\theta \theta }}a^{\theta }.
\label{a_zamo}
\end{equation}%
The scalar square of acceleration%
\begin{equation}
a^{2}=a^{\mu }a_{\mu }=(a_{o}^{(r)})^{2}+(a_{o}^{(\theta
)})^{2}+(a_{o}^{(\varphi )})^{2}-(a_{o}^{(t)})^{2}\text{.}  \label{a2}
\end{equation}

Calculating the Christoffel symbols, one can obtain under assumption of
equatorial motion: 
\begin{equation}
a_{o}^{(r)}=\dfrac{1}{\sqrt{A}}\Bigg\{u^{r}\partial _{r}u^{r}-\dfrac{1}{2}%
\dfrac{\partial _{r}A}{A}\left( u^{r}\right) ^{2}-\dfrac{A}{2}\Big(\mathcal{X%
}^{2}\partial _{r}(N^{-2})-\mathcal{L}^{2}\partial _{r}(g_{\varphi \varphi
}^{-1})-2\dfrac{\mathcal{XL}}{N^{2}}\partial _{r}\omega \Big)\Bigg\},
\label{a_r_o}
\end{equation}%
\begin{equation}
a_{o}^{(t)}=N\Big\{u^{r}\partial _{r}u^{t}+u^{r}\Big(\dfrac{\partial
_{r}N^{2}}{N^{2}}\dfrac{\mathcal{X}}{N^{2}}+\dfrac{\mathcal{L}}{N^{2}}%
\partial _{r}\omega \Big)\Big\},  \label{a_t_o}
\end{equation}%
\begin{equation}
a_{o}^{(\varphi )}=\sqrt{g_{\varphi \varphi }}\Big\{u^{r}(\partial
_{r}u^{\varphi }-\omega \partial _{r}u^{t})-u^{r}\Big(\dfrac{\mathcal{X}}{%
N^{2}}\partial _{r}\omega +\mathcal{L}\partial _{r}(g_{\varphi \varphi
}^{-1})\Big)\Big\}.  \label{a_phi_o}
\end{equation}%
Component $a^{\theta }=0$ because we consider equatorial motion with respect
to which all metric functions are symmetric that causes cancellation of all
terms in $a^{\theta }$.

Expressions for $a_{o}^{(t)}$, $a_{o}^{(r)}$ and $a_{o}^{(\varphi )}$ may be
simplified by the substitution of the expression for the four-velocity (\ref%
{u}): 
\begin{equation}
a_{o}^{(r)}=\mathcal{X}\dfrac{\sqrt{A}}{N^{2}}\Big(\partial _{r}\mathcal{X}+%
\mathcal{L}\partial _{r}\omega -\dfrac{N^{2}}{\mathcal{X}}\dfrac{\mathcal{L}%
\partial _{r}\mathcal{L}}{g_{\varphi \varphi }}\Big),  \label{ra}
\end{equation}%
\begin{equation}
a_{o}^{(t)}=\dfrac{u^{r}}{N}(\partial _{r}\mathcal{X}+\mathcal{L}\partial
_{r}\omega ),  \label{ta}
\end{equation}%
\begin{equation}
a_{o}^{(\varphi )}=\dfrac{u^{r}}{\sqrt{g_{\varphi \varphi }}}\partial _{r}%
\mathcal{L}.  \label{phi}
\end{equation}

Eqs. (\ref{ta}), (\ref{phi}) agree with eqs. (116), (117) of \cite{tz13}.

Equivalently, we can write: 
\begin{equation}
a_{o}^{(r)}=\mathcal{X}\dfrac{\sqrt{A}}{N^{2}}\Big(\partial _{r}(\mathcal{X}+%
\mathcal{L}\omega )-\Big(\omega +\dfrac{N^{2}}{\mathcal{X}}\dfrac{\mathcal{L}%
}{g_{\varphi \varphi }}\Big)\partial _{r}\mathcal{L}\Big),  \label{ar_2}
\end{equation}%
\begin{equation}
a_{o}^{(t)}=\sigma P\dfrac{\sqrt{A}}{N^{2}}\Big(\partial _{r}(\mathcal{X}+%
\mathcal{L}\omega )-\omega \partial _{r}\mathcal{L}\Big),  \label{at_2}
\end{equation}%
\begin{equation}
a_{o}^{(\varphi )}=\sigma P\dfrac{\sqrt{A}\partial _{r}\mathcal{L}}{N\sqrt{%
g_{\varphi \varphi }}}.  \label{aphi_2}
\end{equation}

\section{Fine-tuned particles}

\label{fin-tun}

We are interested in trajectories that are (i) compatible with finite
acceleration near the horizon, (ii)\ lead to an indefinitely large growth of
energy $E_{c.m.}$ due to particle collision there. Property (ii) implies
that the proper time required to reach the horizon is infinite for a
fine-tuned (subcritical, critical or ultracritical) particle, so that it
approaches the horizon only asymptotically and cannot cross it. (For the
Kerr metric this was noticed in \cite{ted}, a general proof will be given
below in Sec. \ref{proper}). Correspondingly, it is the OZAMO frame which is
natural for them (see below for more detail) since a corresponding observer
does not cross the horizon. Therefore, we can write asymptotic expansion for
acceleration near the horizon in the form%
\begin{equation}
a_{o}^{(t)}=(a_{o}^{(t)})_{n_{0}}v^{n_{0}}+o(v^{n_{0}}),~~~a_{o}^{(r)}=(a_{o}^{(r)})_{n_{1}}v^{n_{1}}+o(v^{n_{1}}),~~~a_{o}^{(\varphi )}=(a_{o}^{(\varphi )})_{n_{2}}v^{n_{2}}+o(v^{n_{2}}),
\end{equation}%
where $n_{0},~n_{1},~n_{2}$ should be non-negative. As we mentioned above,
we will proceed in such a way: we set a near-horizon trajectory
(equivalently, numbers $c$ and $b$ that appear in (\ref{c}) and (\ref{Lb}))
and calculate accelerations, thus finding $n_{0}$, $n_{1}$ and $n_{2}$.
Requiring $n_{0}$, $n_{1}$ and $n_{2}$ to be non-negative, we find
physically achievable trajectories that can produce the BSW effect. To
realize this scheme, we have to establish several important restrictions on
the parameters of our system.

One important reservation is in order. If we take into account (\ref{puu}) -
(\ref{pultra}), it follows from (\ref{ar_2}) and (\ref{at_2}) that in the
case of subcritical and critical particles $n_{0}=n_{1}$, while for
ultracritical ones $n_{0}>n_{1}$. Thus $n_{0}\geq n_{1}$ and regularity of $%
a_{o}^{(r)}$ implies regularity of $a_{o}^{(t)}$, so it is sufficient to
analyze $a_{o}^{(r)}$ and $a_{o}^{(\varphi)}$ only.

To find relation between $n_{0}$, $n_{1}$, $n_{2}$ and $c,$ $b$, we express
the four-velocity in terms of quantities $\mathcal{X}$ and $\mathcal{L}$
introduced above.

\subsection{Relationship between acceleration and radial velocity near the
horizon}

\label{beh_acc}

In this subsection we will find explicitly the asymptotic behavior of the
expressions for accelerations listed above. Before proceeding further, we
want to make some important reservations. We are interested in situations,
when the tetrad components of acceleration are finite in a relevant frame.
By this frame, we imply a frame comoving with respect to a particle or any
other one that moves with respect to it with a finite velocity giving a
finite local Lorentz boost between them. For a usual particle the role of
such a frame is played by a frame attached to a free-falling observer
(FZAMO, if for simplicity we choose an observer with a zero angular
momentum). However, in the OZAMO frame, its components may diverge since the
Lorentz boost becomes singular. By contrast, the fine-tuned particles cannot
cross a horizon. Therefore, it is the OZAMO frame which is natural for them,
so that the tetrad components of acceleration in the OZAMO frame should stay
finite. (This general issue is discussed in more detail in Sec. III of \cite%
{tz13}. In particular, see eqs. 70, 71 there.). It is the study of concrete
near-horizon asymptotic expressions that we are now turning to.

Combining eq. (\ref{s}) with $s>0$, (\ref{P}) and (\ref{ur}) and using (\ref%
{c}), we arrive at the following set of cases.

\begin{itemize}
\item $s<p/2$, subcritical particle: In this case $u^{r}=\sigma \sqrt{\dfrac{%
A_{q}}{\kappa _{p}}}\mathcal{X}_{s}v^{s+(q-p)/2}+o(v^{s+(p-q)/2})$ that
gives us relation $s=\dfrac{p-q}{2}+c.$

\item $s=p/2$, critical particle: In this case $P=\sqrt{\mathcal{X}%
_{p/2}^{2}-\kappa _{p}\Big(\dfrac{\mathcal{L}_{H}^2}{g_{\varphi }}+1\Big)}%
\nu ^{p/2}+o(u^{p/2})$,\newline
$u^{r}=\sigma\sqrt{\dfrac{A_{q}}{\kappa _{p}}}\sqrt{\mathcal{X}%
_{p/2}^{2}-\kappa _{p}\Big(\dfrac{\mathcal{L}_{H}^2}{g_{\varphi }}+1\Big)}%
\nu ^{q/2}+o(u^{p/2}) $.

\item $s=p/2$, ultracritical particle: In this case $u^r=(u^r)_c v^c+...$,
thus $P\approx\sqrt{\dfrac{\kappa_p}{A_q}}(u^r)_c v^{\frac{p-q}{2}+c}.$
\end{itemize}

The case $s>p/2$ is impossible, because $P$ would become imaginary. Now let
us compute components of acceleration for each $s$. We will write the
asymptotics for $P$ in the cases of subcritical and critical particles in
the form $P\approx P_{s}\nu ^{s}+o(\nu ^{s})$, where $P_{s}=\mathcal{X}_{s}$
for $s<p/2$ and $P_{s}=\sqrt{\mathcal{X}_{p/2}^{2}-\kappa _{p}\Big(\dfrac{%
\mathcal{L}_{H}^{2}}{g_{\varphi }}+1\Big)}$ for $s=p/2$. For the
ultracritical particle we will write $P\approx P_{\frac{p-q}{2}+c}v^{\frac{%
p-q}{2}+c}$, where $P_{\frac{p-q}{2}+c}=\sqrt{\dfrac{\kappa _{p}}{A_{q}}}%
(u^{r})_{c}.$

Let us analyze behavior of radial acceleration. It follows from (\ref{a_r_o}%
) that%
\begin{equation}
a_{o}^{(r)}\approx \dfrac{1}{\sqrt{A_{q}}}\Bigg\{\dfrac{A_{q}}{\kappa _{p}}%
P_{s}^{2}\Big(s-\dfrac{p}{2}\Big)\nu ^{s}+\dfrac{A_{q}}{\kappa _{p}}\mathcal{%
X}_{s}^{2}\dfrac{p}{2}\nu ^{s}+A_{q}\dfrac{\mathcal{X}_{s}\mathcal{L}_{H}}{%
\kappa _{p}}k\omega _{k}\nu ^{k}\Bigg\}\nu ^{s+q/2-p-1}.  \label{a_r}
\end{equation}%
Note that the term with $\mathcal{L}^{2}$ which is present in (\ref{a_r_o})
is of higher order. To see this, let us consider parentheses in (\ref{a_r_o}%
). The first term is of order of $v^{2s-p-1}$ and, as $s\leq p/2$, this term
is divergent. Meanwhile, the $\mathcal{L}^{2}$ term $=O(1)$ that proves the
aforementioned statement.

The near-horizon behavior of (\ref{a_r}) depends on what is bigger: $s$ or $%
k $. Using eq. (\ref{a_r}), we can thus write: 
\begin{equation}
n_{1}=\min (s,k)+s+\dfrac{q}{2}-p-1.  \label{48}
\end{equation}

In accordance with reservations made above, now we are not interested in
usual particles, so $s>0$, whereas the case $s=0$ is excluded from
consideration.

The case when $a_{o}^{(r)}$ does not include $\partial _{r}\omega $ deserves
separate attention. This may happen if $\omega $ is constant. As a matter of
fact, such a metric is static. In this case we can redefine angular
coordinate $\tilde{\varphi}=\varphi -\omega t$ that will diagonalize metric (%
\ref{metr}), making it explicitly static. Then in (\ref{a_r}) only two first
terms survive that gives us 
\begin{equation}
n_{1}=2s+\dfrac{q}{2}-p-1.  \label{51}
\end{equation}%
Hereafter, we denote this case as $k=0$.

There is also a special case when coefficients in expansion of $\mathcal{X}$
and $\mathcal{L}$ are such that several terms in powers series (which are,
generally speaking, potentially divergent) in the expression for
acceleration cancel each other. Full cancellation happens, for example, for
freely falling particle, for which $\mathcal{X+}\omega \mathcal{L}%
=\varepsilon $, where $\varepsilon $ and $\mathcal{L}$ are constants that
gives us zero acceleration. Then, $\partial _{r}(\mathcal{X+}\omega \mathcal{%
L)}=0$ exactly. In a more general case we can consider 
\begin{equation}
\varepsilon =\mathcal{X}+\omega \mathcal{L}=\varepsilon _{0}+\mathrm{terms~of%
}~v^{m}~\mathrm{order},~m>k.  \label{spec}
\end{equation}

Using relation (\ref{ar_2}) for $a_{o}^{(r)}$, we see that the first term in
brackets has the order $v^{m-1}$, while the second one has the order of $%
v^{b-1}$.

Thus in this case 
\begin{equation}
n_{1}=\min (m,b)+s+\dfrac{q}{2}-p-1.  \label{53}
\end{equation}

However, we will not pay much attention to this case further.

Now we want to rewrite all possible solutions for $n_{1}$ in terms of $c$.
In the case of usual, subcritical and critical particles we can use relation 
$s=\dfrac{p-q}{2}+c$ that gives us 
\begin{equation}
n_{1}=\min \Big(\dfrac{p-q}{2}+c,k\Big)+c-1-\dfrac{p}{2}~\mathrm{if}~k\neq 0,
\end{equation}%
\begin{equation}
n_{1}=2c-1-\dfrac{q}{2}~\mathrm{if}~k=0,
\end{equation}%
\begin{equation}
n_{1}=\min (m,b)+c-\dfrac{p}{2}-1~\mathrm{if}~\mathcal{X}+\omega \mathcal{L}%
=\varepsilon _{0}+O(v^{m}).  \label{58}
\end{equation}

Thus we found the expressions that include $n_{1}$, $c$, $s$, $p,$ $q,$ $k$.
Our goal is to transform them to the form $c=c(n_{1},p,q,k$). Then we take
into account the data from Table \ref{class_of_tr} in combination with the
requirement $n_{1}\geq 0$. This can give us restrictions on metric
parameters relevant for different types of trajectories. Using the above
formulas, we find for subcritical and critical particles 
\begin{equation}
c=%
\begin{cases}
\dfrac{2n_{1}+2+q}{4}~~\mathrm{if}~~n_{1}\leq 2k+\dfrac{q-2-2p}{2}\text{, }%
k\neq 0 \\ 
n_{1}+\dfrac{p}{2}+1-k~~\mathrm{if}~~n_{1}>2k+\dfrac{q-2-2p}{2},\text{ }%
k\neq 0 \\ 
\dfrac{2n_{1}+2+q}{4}~\mathrm{if}~k=0 \\ 
n_{1}+\dfrac{p}{2}+1-\min (m,b)~\mathrm{if}~\mathcal{X}+\omega \mathcal{L}%
=\varepsilon _{0}+O(v^{m})\label{a_rel_2}~\mathrm{where}~m>k%
\end{cases}%
\end{equation}%
where the relation $s=\dfrac{p-q}{2}+c$ was used.

The first solution in (\ref{a_rel_2}) exists for $n_{1}\geq 0$ if 
\begin{equation}
k\geq \dfrac{2p+2-q}{4}  \label{fk}
\end{equation}%
only, while the second one exists for all $k$.

For ultracritical particles we cannot use the aforementioned formula for $s$%
. In this case $c>\dfrac{q}{2},$ $s=\frac{p}{2}$. Then, we obtain from (\ref%
{48}) and (\ref{51}): 
\begin{gather}
c\mathrm{~may~be~any~value~grater~then~}\dfrac{q}{2}, \\
n_{1}=\mathrm{min}\Big(\dfrac{p}{2},k\Big)+\dfrac{q-p}{2}-1~\mathrm{if}%
~k\neq 0, \\
n_{1}=\dfrac{q-2}{2}~\mathrm{if}~k=0.  \label{uc_n1}
\end{gather}

The special case $\mathcal{X}+\omega \mathcal{L}=\varepsilon _{0}+O(v^{m})$
gives us, according to (\ref{53}),

\begin{equation}
n_{1}=\min (m,b)+\dfrac{q-p}{2}-1.
\end{equation}

We remind a reader that, according to what is said above, there is no need
to require the regularity of the time component of acceleration. It is valid
automatically if the radial one is regular.

Now, let us consider $a_{o}^{(\varphi )}$. It follows from (\ref{phi}) and (%
\ref{Lb}) that for any type of particle 
\begin{equation}
a_{o}^{(\varphi )}\approx \mathcal{L}_{h}\nu ^{c+b-1},
\end{equation}%
where $\mathcal{L}_{h}$ is some constant, so 
\begin{equation}
n_{2}=c+b-1\rightarrow b=n_{2}+1-c.  \label{b_rel}
\end{equation}%
As we require $n_{2}\geq 0$, this gives us the restriction $b\geq 1-c$.
According to Table \ref{class_of_tr}, this entails $b\geq 1-\dfrac{q}{2}$
for critical and ultracritical trajectories.

One reservation is in order. The formulas under discussion include also the
case when $a_{o}^{(\varphi )}=0$, so $\mathcal{L}=$const. Then, formally,
one can put $b\rightarrow \infty $. Correspondingly, it drops out from (\ref%
{a_rel_2}).

More information can be extracted from Table \ref{n2_tab}.

\begin{table}[tbp]
\centering
\begin{tabular}{|c||c|c|}
\hline
~ & Type of trajectory & $n_2$ \\ \hline\hline
1 & Usual & $\dfrac{q-p}{2}+b-1$ \\ \hline
2 & Subcritical & $\dfrac{q-p}{2}+b-1<n_{2}<b-1+\frac{q}{2}$ \\ \hline
3 & Critical & $\dfrac{q}{2}+b-1$ \\ \hline
4 & Ultracritical & $n_{2}>\dfrac{q}{2}+b-1$ \\ \hline
\end{tabular}%
\caption{Behaviour of the angular component of acceleration.}
\label{n2_tab}
\end{table}

Now, we will analyze for which subspaces in the space of parameters $(p,q,k)$
we can have near-horizon trajectories with non-negative $n_{0}$, $n_{1}$ and 
$n_{2}$ for all types of fine-tuned particles described in Table \ref%
{class_of_tr}.

\subsection{Subcritical particles}

Let us start with subcritical particles. According to Table \ref{class_of_tr}%
, $\dfrac{q-p}{2}<c<\dfrac{q}{2}$. If we use the first solution in (\ref%
{a_rel_2}), we get $\dfrac{q-2-2p}{2}<n_{1}<\dfrac{q-2}{2}$. Now, let us
find how it correlates with the condition of existence of the first solution
in (\ref{a_rel_2}): $n_{1}\leq 2k+\dfrac{q-2-2p}{2}$. For $k<\dfrac{p}{2},$
the condition $n_{1}\leq 2k+\dfrac{q-2-2p}{2}$ is stronger than $n_{1}<%
\dfrac{q-2}{2}$, for $k\geq \dfrac{p}{2}$ the condition $n_{1}<\dfrac{q-2}{2}
$ becomes stronger. The lower bound for $n_{1}$ is the same for all positive 
$k$. So, we can conclude that for the first solution in \textrm{~}(\ref%
{a_rel_2}) 
\begin{gather}
\dfrac{q-2-2p}{2}<n_{1}~\leq 2k+\dfrac{q-2-2p}{2}~\mathrm{if}~k<\dfrac{p}{2},
\label{subcr_1_k<} \\
\dfrac{q-2-2p}{2}<n_{1}<\dfrac{q-2}{2}~\mathrm{if}~k\geq \dfrac{p}{2}.
\label{subcr_1}
\end{gather}

For the second solution in (\ref{a_rel_2}) the condition $\dfrac{q-p}{2}<c<%
\dfrac{q}{2}$ entails $\dfrac{q}{2}-p-1+k<n_{1}<\dfrac{q-p}{2}-1+k$. The
condition $n_{1}>2k+\dfrac{q-2-2p}{2}$ is stronger than the lower bound.
However, it is weaker then the upper one, if $k<\dfrac{p}{2}$. Thus we can
conclude that for the second solution in (\ref{a_rel_2}) 
\begin{gather}
2k+\dfrac{q-2-2p}{2}\leq n_{1}<~\dfrac{q-p}{2}-1+k\text{ }\mathrm{if}~k<%
\dfrac{p}{2},  \label{subcr_2_k<} \\
\mathrm{Second~solution~is~impossible}~\mathrm{if}~k\geq \dfrac{p}{2}.
\label{subcr_2}
\end{gather}%
The special case $k=0$ gives the condition $\dfrac{q-2-2p}{2}<n_{1}<\dfrac{%
q-2}{2}$. This completes the analysis for the subcritical case.

\subsection{Critical particles}

The first solution in (\ref{a_rel_2}) gives us $c=\dfrac{q}{2}=\dfrac{%
2n_{1}+2+q}{4}\rightarrow n_{1}=\dfrac{q-2}{2}$, so $n_{1}$ is non-negative
if $q\geq 2$. The first solution exists if $n_{1}\leq 2k+\dfrac{q-2-2p}{2}$
that gives us $k\geq \dfrac{p}{2}$. If we deal with the second solution, $%
n_{1}+\dfrac{p}{2}+1-k=\dfrac{q}{2}\rightarrow n_{1}=\dfrac{q-2}{2}+k-\dfrac{%
p}{2}$. The condition $n_{1}\geq 0$ gives us $k\geq \dfrac{p-q}{2}+1$. The
second solution exists if $n_{1}>2k+\dfrac{q-2-2p}{2}$ that entails $k<%
\dfrac{p}{2}$. To conclude, we see that 
\begin{gather}
\mathrm{If}~\dfrac{p-q}{2}+1\leq k<\dfrac{p}{2}\text{, }n_{1}=\dfrac{q-2}{2}%
+k-\dfrac{p}{2}.  \label{cr_cond_k<} \\
\mathrm{If}~k\geq \dfrac{p}{2}~n_{1}=\dfrac{q-2}{2}.  \label{cr_cond}
\end{gather}

The special case $k=0$ gives the same result as the first solution in (\ref%
{a_rel_2}).

\subsection{Ultracritical particles}

When we turn to ultracritical trajectories, the situation is somewhat
different. In this case $c$ is independent of $n_{1}$: $c$ may take any
value greater than $\dfrac{q}{2}$, while $n_{1}=\dfrac{q-2}{2}$ if $k\geq 
\dfrac{p}{2}$, $n_{1}=k+\dfrac{q-p}{2}-1$ if $k<\dfrac{p}{2}$ (see (\ref%
{uc_n1})). It would seem that these relations are the same as in the case of
a critical trajectory (because forces are the same). However, as we
discussed in section \ref{beh}, for ultracritical particles an additional
condition (\ref{xi_cond}) has to hold.

\subsection{Classification of different $k-$regions}

In the above consideration, we dicussed the conditions of existence of
different types of fine-tuned particles. Meanwhile, we are also interested
in more subtle details that concern the relations between types of solutions
enumerated in the lines of eq. (\ref{a_rel_2}) and those of trajectories. To
this end, it is convenient to systematize the obtained results and collect
them in an unified scheme. In a natural way, the space of parameters $%
(p,q,k) $ is split depending on a kind of particle (subcritical, critical,
ultracritical) to different regions in which the radial component of
acceleration is finite. Introducing the values of $k$ that correspond to the
borders of such regions, we obtain four different regions for $k$

\begin{gather}
0<k<\dfrac{p-q}{2}+1,~\mathrm{region}\text{ }\mathrm{I,} \\
\dfrac{p-q}{2}+1\leq k<\dfrac{p+1-q/2}{2},~\mathrm{region}\text{ }\mathrm{II}%
, \\
\dfrac{p+1-q/2}{2}\leq k<\dfrac{p}{2},~\mathrm{~region}\text{ }\mathrm{III},
\label{89} \\
k\geq p/2,~\mathrm{region}\text{ }\mathrm{IV.~}
\end{gather}

They are presented on Fig. \ref{k_range}. Here, according to (\ref{fk}), $k=%
\dfrac{2p+2-q}{4}$ is the minimum possible value for which the first
solution in (\ref{a_rel_2}) exists, $k=\dfrac{p}{2}$ is the maximum possible
value for which subcritical particles given by the second solution in (\ref%
{a_rel_2}) can exist according to (\ref{subcr_2}).

In region IV ($k\geq \dfrac{p}{2}$) we have corresponding conditions for
different trajectories. As follows from (\ref{subcr_1}), if $\dfrac{q-2-2p}{2%
}<n_{1}<\dfrac{q-2}{2}$ , we have a subcritical particle. As follows from (%
\ref{cr_cond}), if $n_{1}=\dfrac{q-2}{2}$, a particle is critical, while if $%
n_{1}=\dfrac{q-2}{2}$ and the condition (\ref{xi_cond}) holds, it is
ultracritical.

In region III we note that $\dfrac{2+2p-q}{4}\leq k<\dfrac{p}{2}$ and the
conditions somewhat change. Following (\ref{subcr_1_k<}) and (\ref%
{subcr_2_k<}) we see that the first solution in (\ref{a_rel_2}) gives a
subcritical particle only for $\dfrac{q-2-2p}{2}<n_{1}\leq 2k+\dfrac{q-2-2p}{%
2}$. The second solution in (\ref{a_rel_2}) gives a subcritical particle
only for $2k+\dfrac{q-2-2p}{2}<n_{1}<\dfrac{q-p}{2}-1+k$ (see (\ref%
{subcr_2_k<})). If we are interested in the type of trajectory only, then
these two regions may be joined. This gives us that for all $\dfrac{q-2-2p}{2%
}<n_{1}<\dfrac{q-p}{2}-1+k$ we have subcritical particles. Critical
particles can be obtained only if $n_{1}=\dfrac{q-2}{2}+k-\dfrac{p}{2}$ (see
(\ref{cr_cond_k<})). In the ultracritical case acceleration is the same as
for the critical case ($n_{1}=\dfrac{q-2}{2}+k-\dfrac{p}{2}$), but condition
(\ref{xi_cond}) has to hold.

Region II gives the same conditions as region III. However, the first
solution in (\ref{a_rel_2}) is absent (as we noted in discussion after (\ref%
{a_rel_2})). Thus we can get subcritical particles only for $2k+\dfrac{q-2-2p%
}{2}<n_{1}<\dfrac{q-p}{2}-1+k$ in II region (where $c=n_{1}+1+p/2-k$).
Relations for critical and ultracritical are the same as for region III.

In region I all types of trajectories give negative $n_1$, because for all
trajectories $n_1\leq k+\dfrac{q-p}{2}-1$, but as $k<\dfrac{p-q}{2}+1$ in
this region, we see that $n_1<0$ for all types of trajectories.

All these results are summarized in Table \ref{reg_tab} and Fig.\ref{k_range}%
. We added also special cases there: $k=0$. As we see, the restrictions on $%
n_{1}$ and/or their exact expressions look different in different $k-$%
regions.

To complete the picture of all possible cases, we explicitly present tables
for different horizons. In the case of a non-extremal horizon ($q<2$) all
trajectories experience infinite forces (see Table \ref{tab_q<2}). In the
case of extremal horizons ($q=2$) a finite force acts only if particle is
critical or ultracritical. In the second case it is possible only if $k=0$
or it lies in III and IV region (see Table \ref{tab_q=2}). In the case of
ultracritical horizon ($q>2$) the situation is more complicated and all
possible trajectories are listed in Table \ref{reg_tab}.

{\centering
\begin{figure}[tbp]
\includegraphics[width=17truecm]{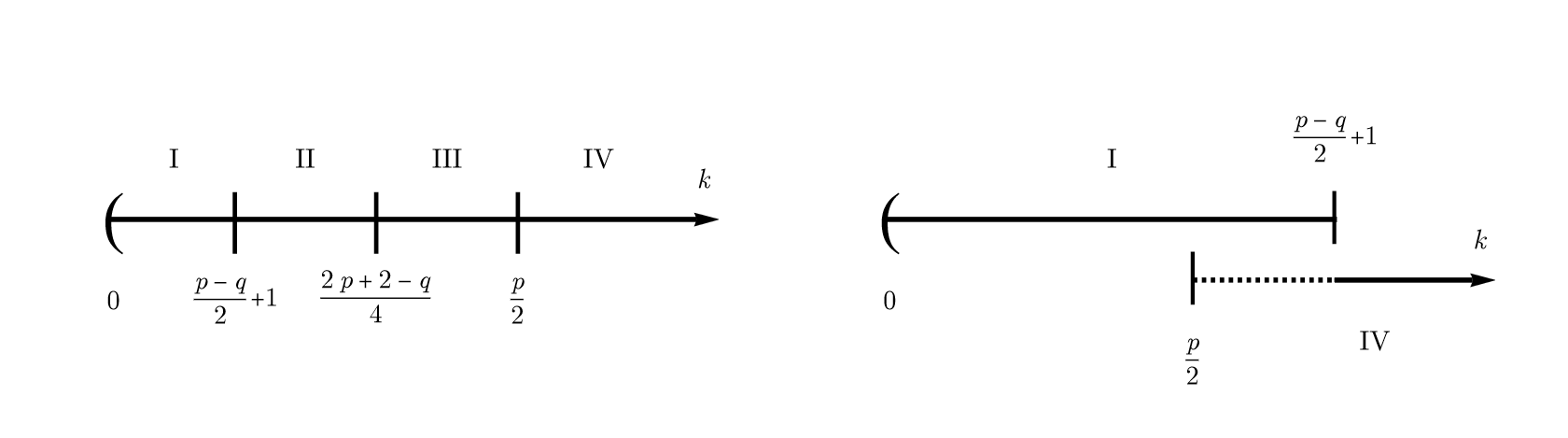}
\caption{Relevant regions of $k$. Left panel corresponds to $q> 2 $. Right
panel corresponds to $q\leq 2$, where regions II and III do not exist
anymore and also points at regions I and IV overlap. We define points on
intersection to belong to region I (for explanation see subsection \protect
\ref{class-of-kreg}).}
\label{k_range}
\end{figure}
}

\begin{table}[]
\centering
{\small 
\begin{tabular}{|c||c|c|c|c|}
\hline
~ & {$k$ Region} & {$n_1$ Range} & $c$ & {Type of trajectory} \\ \hline\hline
~ & \multicolumn{4}{|c|}{{\large {Stationary metric}}} \\ \hline
1 & I & \multicolumn{3}{|c|}{For any type of trajectory $n_1$ is negative
(forces diverge)} \\ \hline
\multirow{3}{*}{2} & \multirow{3}{*}{II} & $2k+\dfrac{q-2-2p}{2}<n_1<\dfrac{%
q-p}{2}-1+k$ & $n_1+1+p/2-k$ & {Subcritical} \\ 
~ & ~ & $n_1=\dfrac{q-p}{2}-1+k$ & $q/2$ & {Critical} \\ 
~ & ~ & $n_1=\dfrac{q-p}{2}-1+k$ and (\ref{xi_cond}) & Any $c>\dfrac{q}{2}$
& {Ultracritical} \\ \hline
\multirow{4}{*}{3} & \multirow{4}{*}{III} & $\max\Big(0,\dfrac{q-2-2p}{2}%
\Big)< n_1\leq 2k+\dfrac{q-2-2p}{2}$ & $\dfrac{2n_1+2+q}{4}$ & {Subcritical}
\\ 
~ & ~ & $2k+\dfrac{q-2-2p}{2}<n_1<\dfrac{q-p}{2}-1+k$ & $n_1+1+p/2-k$ & {%
Subcritical} \\ 
~ & ~ & $n_1=\dfrac{q-p}{2}-1+k$ & $q/2$ & {Critical} \\ 
~ & ~ & $n_1=\dfrac{q-p}{2}-1+k$ and (\ref{xi_cond}) & Any $c>\dfrac{q}{2}$
& {Ultracritical} \\ \hline
\multirow{3}{*}{4} & \multirow{3}{*}{IV} & $\max\Big(0,\dfrac{q-2-2p}{2}\Big)%
< n_1 <\dfrac{q-2}{2}$ & $\dfrac{2n_1+2+q}{4}$ & {Subcritical} \\ 
~ & ~ & $n_1=\dfrac{q-2}{2}$ & $q/2$ & {Critical} \\ 
~ & ~ & $n_1 =\dfrac{q-2}{2}$ and (\ref{xi_cond}) & Any $c>\dfrac{q}{2}$ & {%
Ultracritical} \\ \hline
~ & \multicolumn{4}{|c|}{{\large {Static metric}}} \\ \hline
5 & $k=0$ & \multicolumn{3}{|c|}{Same results as in IV for stationary metric}
\\ \hline
\end{tabular}
}
\caption{Classification of near-horizon trajectories for different $k$%
-regions for $q> 2$ (ultraextremal horizon). The 4-th solution in (\protect
\ref{a_rel_2}) is not presented in this table. Definitions of different $k-$%
regions are given on FIG. \protect\ref{k_range}.}
\label{reg_tab}
\end{table}

\begin{table}[]
\centering
\begin{tabular}{|c||c|c|c|c|}
\hline
~ & {$k$ Region} & {$n_1$ Range} & $c$ & {Type of trajectory} \\ \hline\hline
~ & \multicolumn{4}{|c|}{{\large {Stationary metric}}} \\ \hline
1 & I & \multicolumn{3}{|c|}{For any type of trajectory $n_1$ is negative
(forces diverge)} \\ \hline
\multirow{2}{*}{2} & \multirow{2}{*}{IV} & $n_1=0$ & 1 & {Critical} \\ 
~ & ~ & $n_1=0$ and (\ref{xi_cond}) & Any $c>1$ & {Ultracritical} \\ \hline
~ & \multicolumn{4}{|c|}{{\large {Static metric}}} \\ \hline
3 & $k=0$ & \multicolumn{3}{|c|}{Same results as in IV for stationary metric}
\\ \hline
\end{tabular}%
\caption{Classification of near-horizon trajectories for different $k$%
-regions for $q=2$ (extemal horizon). Regions II and III in this case are
absent, so they are not presented in this table. The 4-th solution in (%
\protect\ref{a_rel_2}) is also not presented in this table. Definitions of
different $k-$regions are given on FIG. \protect\ref{k_range}.}
\label{tab_q=2}
\end{table}

\begin{table}[]
\centering
\begin{tabular}{|c||c|c|c|c|}
\hline
~ & {$k$ Region} & {$n_1$ Range} & $c$ & {Type of trajectory} \\ \hline\hline
~ & \multicolumn{4}{|c|}{{\large {Stationary metric}}} \\ \hline
1 & I and IV & \multicolumn{3}{|c|}{For any type of trajectory $n_1$ is
negative (forces diverge)} \\ \hline
~ & \multicolumn{4}{|c|}{{\large {Static metric}}} \\ \hline
2 & $k=0$ & \multicolumn{3}{|c|}{For any type of trajectory $n_1$ is
negative (forces diverge)} \\ \hline
\end{tabular}%
\caption{Classification of near-horizon trajectories for different $k$%
-regions for $q< 2$ (non-extremal horizon). Regions II and III in this case
are absent, so they are not presented in this table. The 4-th solution in (%
\protect\ref{a_rel_2}) is also not presented in this table. Definitions of
different $k-$regions are given on FIG. \protect\ref{k_range}.}
\label{tab_q<2}
\end{table}

\label{class-of-kreg}

However, there are several possibilities for these regions to intersect or
disappear. Let us start with I region. From definition, we see that it
disappears for $q\geq p+2$, and then the whole range consists only from II,
III and IV regions. Region II is absent for $q\leq 2$. However, if $q\geq
2p+2$, the upper bound becomes negative and then this region also
disappears, so that the whole range consists of III and IV regions. Region
III disappears only if $q\leq 2$. In this case only regions I and IV remain.
Moreover, they start to intersect in this case. For definiteness, we
prescribe the corresponding points on this intersection to region I.
According to Table \ref{tab_q<2}, this choice is not important because in
both I and IV regions $n_{1}$ is negative.

All obtained above results are collected in Table \ref{posibl_rangs} and
represented on FIG. \ref{diag_fin_2}. On this figure, we show cross-section
of 3-parametric space: $k,~p$ and $q$ by plane $k=const$ (FIG. \ref%
{diag_fin_2}). Here, a blue color represents I region, green -II,
orange-III, gray-IV.

\begin{table}[]
\centering
\begin{tabular}{|c|c|}
\hline
Condition for $q$ & Possible $k-$regions \\ \hline\hline
$q\leq 2$ & I and IV \\ \hline
$2< q< p+2$ & I, II, III and IV \\ \hline
$p+2\leq q<2p+2$ & II, III and IV \\ \hline
$q\geq 2p+2$ & III and IV \\ \hline
\end{tabular}%
\caption{Possible $k-$regions depending on $p$ and $q$. Definitions of
different $k-$regions are given on FIG. \protect\ref{k_range}.}
\label{posibl_rangs}
\end{table}

{\centering
\begin{figure}[tbp]
\centering
\includegraphics[width=17truecm]{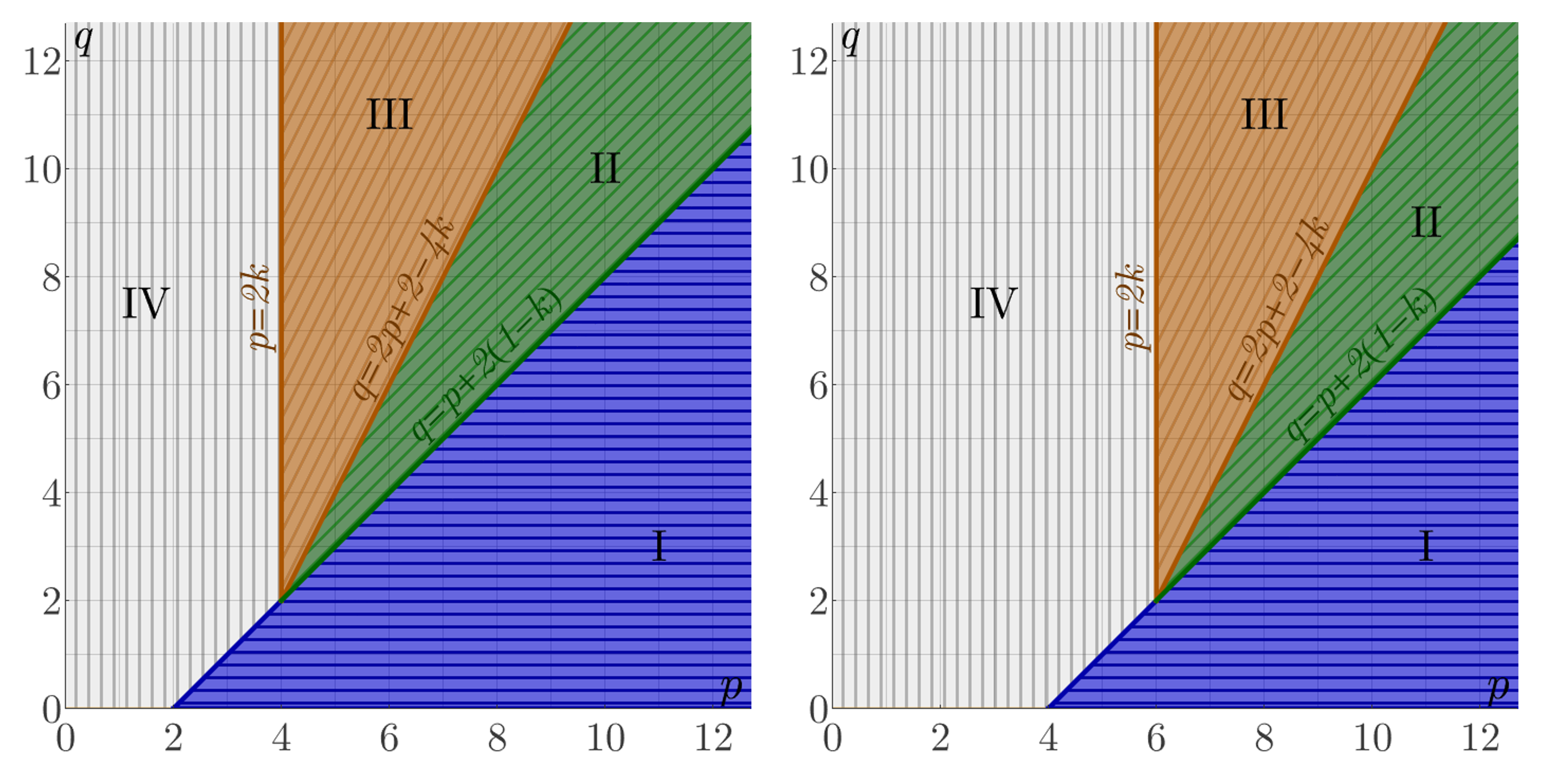}
\caption{Plot showing in which region lies $k$ for all types of horizons,
depending on $p$ and $q$. Left panel is drawn for $k=2$, right panel for $%
k=3 $. Different $k-$regions are represented with different color: IV
(gray), III (orange), II (green), I (blue). }
\label{diag_fin_2}
\end{figure}
}

\subsection{Case $q=p$}

This case has to be considered separately, because of practical importance.
In this case if $0<k<1$, we are at I region, If $1\leq k<\dfrac{p+2}{4}$-in
II, if $\dfrac{p+2}{4}\leq k< \dfrac{p}{2}$-in III, if $\dfrac{p}{2}\leq k$%
-in IV. As in general case, in region I force diverges for all types of
horizons. In region II, as we can see from Table \ref{reg_tab}, all three
types of trajectories are possible. The same holds for III region, but
depending on $k$ expressions for $c$ may be different. Also as in general
case regions II and III are absent if $p<2$

So, summarizing, we have such trajectories in regions II and III:

\begin{itemize}
\item If $0<n_1\leq 2k-1-\dfrac{p}{2}$, trajectory is subcritical and $c=%
\dfrac{2n_1+2+p}{4},$

\item If $2k-1-\dfrac{p}{2}<n_1<k-1$, trajectory is subcritical and $%
c=n_1+1+p/2-k,$

\item If $n_1=k-1$, trajectory is critical,

\item If $n_1=k-1$ and condition (\ref{xi_cond}) holds, trajectory is
ultracritical.
\end{itemize}

In region IV, however, we have:

\begin{itemize}
\item If $0<n_1<\dfrac{p}{2}-1$, trajectory is subcritical,

\item If $n_1=\dfrac{p}{2}-1$, trajectory is critical,

\item If $n_{1}=\dfrac{p}{2}-1$ and condition (\ref{xi_cond}) holds,
trajectory is ultracritical.
\end{itemize}

\section{Results for nonextremal, extremal and ultraextremal horizons}

\label{results_typ}

In previous sections we have analyzed relations between a type of trajectory
and behavior of force depending on characterstics of a horizon. From these
results, we extract now information about possiblity to have finite forces
near the horizon for each type of horizon separately. We also collect the
cases when this is consistent with the BSW effect.

\subsubsection{Non-extremal horizon}

For a non-extremal horizon $p=q=1$. All results corresponding to this case
may be found in Table \ref{tab_q<2}. For subcritical, critical and
ultracritical particles acceleration diverges independently of $k$.

\subsubsection{Extremal horizon}

For extremal horizons $q=2$, while $p$ may take any value $p\geq 2$. All
results corresponding to this case may be found in Table \ref{tab_q=2}. For
subcritical, critical and ultracritical particles acceleration diverges for
all $k<\dfrac{p}{2}$. However if $k\geq \dfrac{p}{2}$ or $k=0$ (static
metric), finite acceleration becomes possible for critical and ultracritical
particles. In doing so, the BSW effect is also possible in the scenario that
corresponds to line 4 in Table \ref{gamma_table}.

\subsubsection{Ultraextremal horizon}

In this case $q,$ $p>2$. All results corresponding to this case may be found
in Table \ref{reg_tab}. In this case for all $k<\dfrac{p-q}{2}+1$
acceleration diverges for all subcritical, critical and ultracritical
trajectories. If $k\geq \dfrac{p-q}{2}+1$ or $k=0$ (static metric),
subcritical, critical and ultracritical trajectories with finite
accelerations can exist. The details of relations between behaviour of the
four-velocity and acceleration can be found in Table \ref{reg_tab}. In doing
so, the BSW effect is described by lines 2, 3 and 4 in Table \ref%
{gamma_table}

The above results are summarized in Table \ref{tab_kregs}.

\begin{table}[tbp]
\centering
\begin{tabular}{|c||c|c|c|}
\hline
~ & {Type of horizon} & {Type of trajectory} & {Region of $k$} \\ 
\hline\hline
\multirow{1}{*}{1} & \multirow{1}{*}{Non-extremal} & All types & %
\multirow{1}{*}{Absent} \\ \hline
\multirow{3}{*}{2} & \multirow{3}{*}{Extremal} & Subcritical & Absent \\ 
\cline{3-4}
~ & ~ & Critical & \multirow{2}{*}{$k\geq p/2$ or $k=0$} \\ 
~ & ~ & Ultracritical & ~ \\ \hline
\multirow{3}{*}{3} & \multirow{3}{*}{Ultraextremal} & Subcritical & %
\multirow{3}{*}{$k\geq \dfrac{p-q}{2}+1$ or $k=0$} \\ 
~ & ~ & Critical & ~ \\ 
~ & ~ & Ultracritical & ~ \\ \hline
\end{tabular}%
\caption{Classification of cases when forces are finite for different types
of horizons and trajectories.}
\label{tab_kregs}
\end{table}

Thus, as long as we are interested in the existence of finite acceleration,
it is sufficient to use the Table \ref{tab_kregs} under discussion only.
However, previous tables give us not only the conditions of such existence
but also much more detailed information about possible rates $n_{i}$ that
characterize the behavior of different components of acceleration.

\section{Particles with finite proper time: kinematic censorship preserved 
\label{proper}}

In this section we prove an interesting consequence of previous results: 
\textit{all non-usual trajectories with a finite proper time either
experience infinite force or the horizon fails to be regular}.

To prove this, we recall that the proper time is finite if $q<p+2$ and the
trajectory satisfies the condition 
\begin{equation}
\dfrac{q-p}{2}\leq c<1  \label{c_rang}
\end{equation}%
(see section \ref{beh}). Now our task is to find, which values of $n_{1}$
can be obtained for such trajectories.

We start with the simplest case $q<2$. As we concluded in Table \ref{tab_q<2}%
, in this case a force is infinite for all non-usual trajectories.

Now let us move to the $q\geq 2$ case. The structure of possible solutions
is more complicated, so we will analyze separately each of solutions
obtained in (\ref{a_rel_2}) under assumption $q\geq 2$.

Let us start with the first solution in (\ref{a_rel_2}). From (\ref{c_rang})
one can obtain 
\begin{equation}
\dfrac{q-2-2p}{2}\leq n_{1}<\dfrac{2-q}{2}.
\end{equation}

As $2\leq q<p+2$, both lower and upper bounds are negative that gives us
negative $n_{1}$. Also note that the third solution in (\ref{a_rel_2}) gives
the same result.

From the second solution of (\ref{a_rel_2}) one can obtain 
\begin{equation}
\dfrac{q-2-2p}{2}+k<n_{1}<k-\dfrac{p}{2}.  \label{2_tau}
\end{equation}

It would seem that this can give positive $n_{1}$. However, we will show
that this is impossible. Let us start with subcritical particles. For them,
the second solution in (\ref{a_rel_2}) exists only if $k<\dfrac{p}{2}$ (see (%
\ref{subcr_2_k<})). As $q<p+2$, both lower and upper bounds in (\ref{2_tau})
are negative. For critical and ultracritical particles we have $c\geq \dfrac{%
q}{2}$. However, (\ref{c_rang}) requires $c<1$ that leads to $q<2$ for which
case forces diverge, as is pointed out above.

The forth solution gives us

\begin{equation}
\min(m,b)+\dfrac{q}{2}-p-1<n_1<\min(m,b)-\dfrac{p}{2}.
\end{equation}

By itself, this inequality can give us non-negative $n_{1}$, provided (\ref%
{spec}) holds with $m\geq p/2$ and $b\geq p/2$. However, this case is
impossible because of another reason, not connected with a force. Namely, we
can obtain a finite proper time for non-regular horizons only. Indeed,
non-usual trajectories can exist if $\varepsilon _{0}=\omega _{H}L_{H}$ only
that gives us $\mathcal{X}\sim -\hat{\omega}_{k}L_{H}v^{k}$. Thus $s=k$.
Meanwhile, a finite proper time can be obtained for $c<1$ only. For
subcritical particles (for which relation $c=\dfrac{q-p}{2}+k$ holds) this
entails $k<\dfrac{p-q}{2}+1$. But this is not consistent with the regularity
condition (\ref{k}).

In the case of critical and ultracritical particles $c\geq \dfrac{q}{2}$. At
the beginning of section \ref{beh_acc} we noted that the requirement $%
P^{2}\geq 0$ entails $s\leq \dfrac{p}{2}$ that implies $k\leq \dfrac{p}{2}$,
so we can write $c\geq \dfrac{q-p}{2}+k$. As a proper time is finite only if 
$c<1$, this also gives us violation of regularity condition (\ref{k}).

This completes the proof of our initial statement.

\section{Usual particles\label{us}}

As we already said, in this investigation we mainly investigate non-usual
particles since this is an essential ingredient of the BSW effect.
Meanwhile, consideration of near-horizon trajectories of usual particles can
be also of some interest beyond the context of the BSW effect. In general,
the time and radial components of acceleration for such particles diverge
near the horizon in the OZAMO\ frame (see subsection \ref{beh_acc} above).
More precisely, the asymptotic form of the acceleration near the horizon
takes the form (see eqs. (70) and (71) in \cite{tz13})

\begin{equation}
a_{o}^{(r)}=\dfrac{(a_{f}^{(t)})_{0}-(a_{f}^{(r)})_{0}}{N}%
+[(a_{f}^{(t)})_{1}-(a_{f}^{(r)})_{1}]+...
\end{equation}%
\begin{equation}
a_{o}^{(t)}=-\dfrac{(a_{f}^{(t)})_{0}-(a_{f}^{(r)})_{0}}{N}%
-[(a_{f}^{(t)})_{1}-(a_{f}^{(r)})_{1}]+...
\end{equation}%
where 
\begin{equation}
a_{f}^{(t,r)}=(a_{f}^{(r,t)})_{0}+(a_{f}^{(t,r)})_{1}N+...
\end{equation}%
Only in the exceptional case when $(a_{f}^{(t)})_{0}=(a_{f}^{(r)})_{0}$, $%
a_{o}^{(t,r)}$ remains finite. It is worth noting that, as the
four-acceleration is a space-like vector, it follows from (\ref{a2}) that in
this case at least one of its angular components should be nonzero.

We can pose a question, when such a case can be realized. In our approach
this means the condition $n_{i}\geq 0$. For usual particles for all types of
horizons the expressions for $n_{i}$ are given by (\ref{n1_u}), (\ref{n0_u})
and (\ref{n2_u}) (for stationary case) and by (\ref{n1_u_k0}), (\ref{n0_u_k0}%
) and (\ref{n2_u}).

Let us start with the radial component of acceleration. Usual particles
correspond to $s=0$ or, equivalently, $c=\dfrac{q-p}{2}$. In this case we
cannot use (\ref{a_r}), because $v^{s}$ terms in (\ref{a_r}) vanish. To
analyze higher order terms which become dominant in this case, we use eq. (%
\ref{ra}) and substitute expansion for $\mathcal{X}$ in the form 
\begin{equation}
\mathcal{X}=\mathcal{X}_{0}+\mathcal{X}_{s_{1}}v^{s_{1}}+o(v^{s_{1}}),
\label{a_r_x}
\end{equation}%
where $s_{1}>0$.

Using expression (\ref{ra}), we see that the first term in brackets is $\sim
v^{s_{1}-1}$, the second term is $\sim v^{k-1}$, while the third one is $%
\sim v^{p+b-1}$. This gives us

\begin{equation}
n_{1}=\min (s_{1},k,p+b)+\dfrac{q}{2}-p-1~\mathrm{for~stationary~metric.}
\label{n1_u}
\end{equation}%
In case of static metric $\omega =\mathrm{const}$ and $v^{k}$ terms in (\ref%
{ra}) are absent that gives us 
\begin{equation}
n_{1}=\min (s_{1},p+b)+\dfrac{q}{2}-p-1~\mathrm{for~static~metric.}
\label{n1_u_k0}
\end{equation}

Analyzing the time component, we are faced with the same issue of vanishing
of $v^{s}$ terms. Then, we need to find higher order terms in (\ref{ta}).
Now, in contrast to (\ref{ra}), there is no term with $\partial _{r}\mathcal{%
L}$, so $b$ drops out from the formulas and the conditions analogous to (\ref%
{n1_u}) and (\ref{n1_u_k0}) read 
\begin{gather}
n_{0}=\min (s_{1},k)+\dfrac{q}{2}-p-1~\mathrm{if}~\mathrm{%
for~stationary~metric},  \label{n0_u} \\
n_{0}=s_{1}+\dfrac{q}{2}-p-1~\mathrm{for~static~metric.}  \label{n0_u_k0}
\end{gather}%
The angular component of acceleration can be obtained by substitution $c=%
\dfrac{q-p}{2}$ in (\ref{b_rel}) that gives us 
\begin{equation}
n_{2}=\dfrac{q-p}{2}+b-1.  \label{n2_u}
\end{equation}

The condition of regularity requires 
\begin{equation}
b\geq 1-\dfrac{q-p}{2}.  \label{reg}
\end{equation}

It is worth stressing that the condition of regularity has now different
status for (i) the time and radial components and (ii) the angular one. For
the reason explained above, condition (i) singles out special subclass of
trajectories for which $(a_{f}^{(t)})_{o}=(a_{f}^{(r)})_{o}$. In general
case, (i) can be violated for usual particles. This is because of singular
nature of local Lorentz transformation near the horizon between FZAMO and
OZAMO. Meanwhile, this transformation does not touch upon the component $%
a_{f}^{(\varphi )}$. Therefore, condition (\ref{reg}) is mandatory for
physically acceptable trajectories of usual particles.

\section{Checking results: electromagnetic force}

\label{check-res}

In this section we check our results in the case when a force has
electromagnetic nature using an exact solution of Einstein-Maxwell
equations. To this end, we consider an electrically charged black hole.
Axial symmetry requires that the vector potential has the form 
\begin{equation}
\mathcal{A}=\mathcal{A}_{t}dt+\mathcal{A}_{\varphi }d\varphi .
\end{equation}

As before, we consider axially symmetric metrics (\ref{metr}) and assume the
same symmetry for the electromagnetic field. Then, we introduce generalized
momenta $P_{\mu }$ in a standard way,%
\begin{equation}
p_{\mu }=P_{\mu }-eA_{\mu }\text{,}
\end{equation}

where $e$ is a particle's charge, $p_{\mu }$ being the kinematic momentum
that obeys the normalization condition

\begin{equation}
-m^{2}=g^{\mu \nu }p_{\mu }p_{\nu }.
\end{equation}

Because of symmetry of this system, the quantities $\tilde{E}=-P_{t}$ and $%
\tilde{L}=P_{\varphi }$ remain constant. Thus the normalization condition
for the momentum gives us

\begin{equation}
-m^{2}=-\dfrac{1}{N^{2}}(-\tilde{E}-e\mathcal{A}_{t})^{2}-\dfrac{2\omega }{%
N^{2}}(-\tilde{E}-e\mathcal{A}_{t})(\tilde{L}-e\mathcal{A}_{\varphi })-\Big(%
\dfrac{\omega ^{2}}{N^{2}}-\dfrac{1}{g_{\varphi \varphi }}\Big)(\tilde{L}-e%
\mathcal{A}_{\varphi })^{2}+A(p_{r})^{2}.
\end{equation}%
Introducing $E=\tilde{E}+eA_{t}$, $\mathcal{L}=\dfrac{\tilde{L}}{m}-\dfrac{e%
}{m}A_{\varphi }$ and $\mathcal{X}=\dfrac{E-\omega L}{m}$, we can express $%
u^{r}=A\dfrac{p_{r}}{m}$ in a form: 
\begin{equation}
(u^{r})^{2}=\dfrac{A}{N^{2}}\Big(\mathcal{X}^{2}-N^{2}\Big(1+\dfrac{\mathcal{%
L}^{2}}{g_{\varphi \varphi }}\Big)\Big).
\end{equation}

\begin{itemize}
\item If $\tilde{E}\neq \omega _{H}(\tilde{L}-eA_{\varphi
})_{H}-e(A_{t})_{H} $, $\mathcal{X}_{H}\neq 0$, so $s=0$ that gives us a
usual trajectory.

\item If $\tilde{E}=\omega _{H}(\tilde{L}-eA_{\varphi })_{H}-e(A_{t})_{H}$, $%
\mathcal{X}_{H}=0$, so $s>0$ that gives us a non-usual trajectory.
\end{itemize}

Now let us check what conditions we get in the case the
Kerr-Newman-(anti-)de Sitter space-time.

\subsection{Motion in Kerr-Newman-(anti-)de-Sitter spacetime}

To test our results in some specific case, we use Kerr-Newman-(anti-)de
Sitter solution. In the Boyer-Lindquist coordinates it has a form (see p.
209-210 in \cite{podol}): 
\begin{equation}
ds^{2}=-\dfrac{\Delta _{r}}{\Xi ^{2}\rho ^{2}}\Big(dt-a\sin ^{2}\theta
d\varphi \Big)^{2}+\dfrac{\varrho ^{2}}{\Delta _{r}}dr^{2}+\dfrac{\varrho
^{2}}{\Delta _{\theta }}d\theta ^{2}+\dfrac{\Delta _{\theta }\sin ^{2}{%
\theta }}{\Xi ^{2}\rho ^{2}}\Big(adt-(r^{2}+a^{2})d\varphi \Big)^{2},
\label{kds}
\end{equation}%
\begin{eqnarray}
\rho ^{2} &=&r^{2}+a^{2}\cos ^{2}\theta ,  \label{delta_r} \\
\Delta _{r} &=&(r^{2}+a^{2})(1-\frac{1}{3}\Lambda r^{2})-2Mr+Q^{2}, \\
\Delta _{\theta } &=&1+\frac{1}{3}\Lambda a^{2}\cos ^{2}\theta , \\
\Xi &=&1+\frac{1}{3}\Lambda a^{2},
\end{eqnarray}%
with the vector potential 
\begin{equation}
\mathcal{A}=-\dfrac{Qr}{\Xi \rho ^{2}}(dt-a\sin ^{2}\theta d\varphi
)\implies \mathcal{A}_{t}=-\dfrac{Qr}{\Xi \rho ^{2}},~\mathcal{A}_{\varphi
}=-a\sin ^{2}\theta A_{t}.
\end{equation}

Here $\Lambda $ is a cosmological constant, $a$ is the Kerr parameter, $M$
being the mass of black hole, $Q$ its charge. Using this solution, we can
write: 
\begin{gather}
N^{2}=\dfrac{\Delta _{\theta }\Delta _{r}\rho ^{2}}{\Delta _{\theta
}(r^{2}+a^{2})^{2}-\Delta _{r}a^{2}\sin ^{2}\theta }, \\
\omega =\dfrac{\Delta _{\theta }a(r^{2}+a^{2})-\Delta _{r}a}{\Delta _{\theta
}(r^{2}+a^{2})^{2}-\Delta _{r}a^{2}\sin ^{2}\theta }, \\
A=\dfrac{\Delta _{r}}{\rho ^{2}}.
\end{gather}%
Now 
\begin{equation}
\mathcal{X}=\tilde{\mathcal{X}}+\dfrac{e}{m}(1-a\omega \sin ^{2}{\theta })%
\mathcal{A}_{t},  \label{xi_charg}
\end{equation}%
where $\tilde{\mathcal{X}}=\dfrac{\tilde{E}-\omega \tilde{L}}{m}$.

The horizons correspond to zeros of $\Delta _{r}$ function. Then we see that
near the horizon $A\approx N^{2}\approx \Delta _{r}$. The same behavior of $%
A $ and $N^{2}$ near the horizon entails $q=p$ and this number is equal to
degeneracy of a horizon. In what follows, we consider motion within the
equatorial plane $\theta =\frac{\pi }{2}$ only.

\subsubsection{Non-extremal horizon}

\label{non-extr}

Firstly let us consider a non-extremal horizon. In this case $\Delta
_{r}\approx \alpha v$, where $\alpha $ is a constant. Expanding (\ref%
{xi_charg}) and taking into account that $\theta =\pi /2$ we have%
\begin{equation}
\mathcal{X}=\mathcal{\tilde{X}}_{H}+\dfrac{e}{m}(\mathcal{A}%
_{t})_{H}(1-a\omega _{H})+O(v).  \label{xi_exp}
\end{equation}

If $\tilde{E}=\omega _{H}\tilde{L}+e(A_{t})_{H}(1-a\omega _{H})$, then $%
\mathcal{X}\sim v,~s=1$. However, a corresponding particle cannot reach the
horizon because, as follows from Table \ref{class_of_tr}, $s$ has to satisfy 
$0<s\leq \dfrac{p}{2}$ that in our case gives $0<s\leq \dfrac{1}{2}$.

While if $\tilde{E}\neq \omega _{H}\tilde{L}+e(A_{t})_{H}(1-a\omega _{H})$, $%
\mathcal{X}$ is not zero on horizon that gives us $s=0$ (a usual
trajectory). Using (\ref{xi_exp}) we deduce that $s_{1}=1$. According to (%
\ref{n1_u}) that gives us $n_{1}=\dfrac{q}{2}-p=-\dfrac{1}{2}$. We see that
in this case acceleration in OZAMO frame diverges.

Thus we conclude that for non-extremal horizons electrogeodesics in the
Kerr-Newman-(anti-)de Sitter space-time the existence of critical particles
is forbidden while usual ones experience infinite acceleration in the OZAMO
frame.

\subsubsection{Extremal horizon}

\label{extr}

In the case of extremal horizon $q=p=2$. Expansion for $\mathcal{X}$ has the
similar structure: 
\begin{equation}
\mathcal{X}=\mathcal{\tilde{X}}_{H}-\dfrac{e}{m}(\mathcal{A}%
_{t})_{H}(1-a\omega _{H})+O(v).
\end{equation}

As in the previous case, if $\tilde{E}=\omega _{H}\tilde{L}%
+e(A_{t})_{H}(1-a\omega _{H})$, then $\mathcal{X}\sim v,~s=1$. Such the
near-horizon behavior of $\mathcal{X}$ is allowed, because now $s=\dfrac{p}{2%
}$ that gives us the critical particle. Also note that $k=1$. As in this
case $k=p/2$, we are in region IV. According to Table \ref{tab_q=2}, this
gives us the critical trajectory (2-nd line). As follows from (\ref%
{cr_cond_k<}), in this case $n_{1}=\dfrac{q-2}{2}$. As now $q=2$, we obtain $%
n_{1}=0$, thus such a particle experiences an action of a finite force.

If $\tilde{E}\neq \omega _{H}\tilde{L}+e(A_{t})_{H}(1-a\omega _{H})$, $%
\mathcal{X}_{H}\neq 0$ that gives us $s=0$. From (\ref{xi_exp}) we derive
that $s_{1}=1$. In this case using (\ref{n1_u}) we obtain $n_{1}=-1$. Thus
acceleration is divergent. Thus for extremal horizons only critical
electrogeodesics can reach the horizon with a finite force in the OZAMO
frame.

\subsubsection{Ultraextremal horizon}

\label{ultr-extr}

In this case the horizon is triple, $p=q=3$. We can write the radial
function $\Delta _{r}$ in the form 
\begin{equation}
\Delta _{r}=-\dfrac{\Lambda }{3}(r-b)^{3}(r+r_{0}).
\end{equation}

Comparing this with (\ref{delta_r}), we see that such factorization is
possible only if 
\begin{equation}
b=\dfrac{1}{\sqrt{2\Lambda }}\sqrt{1-\dfrac{\Lambda a^{2}}{3}},~~~~r_{0}=%
\dfrac{3}{\sqrt{2\Lambda }}\sqrt{1-\dfrac{\Lambda a^{2}}{3}},
\end{equation}%
with additional restriction on parameters: 
\begin{equation}
Q^{2}=\dfrac{\Lambda }{3}b^{3}r_{0},~~~M=\dfrac{\Lambda }{2}\Big(b^{2}r_{0}-%
\dfrac{b^{3}}{3}\Big).
\end{equation}%
As $Q$ and $M$ have to be positive, we have the condition $\Lambda >0$. This
restricts us by the the Kerr-Newman-(anti-)de Sitter space-time. Also note
that the triple horizon is cosmological one. However, this does not
influence the behavior of accelerations which we discuss in this section.

In this case expansion for $\mathcal{X}$ is the same: 
\begin{equation}
\mathcal{X}=\dfrac{\tilde{E}-\omega _{H}\tilde{L}}{m}-\dfrac{e}{m}(\mathcal{A%
}_{t})_{H}(1-a\omega _{H})+O(v).
\end{equation}

If $\tilde{E}=\omega _{H}\tilde{L}+e(A_{t})_{H}(1-a\omega _{H})$, then $%
\mathcal{X}\sim v,~s=1$. This value of $s$ gives us subcritical particles,
because the value $s=1$ is lower than $p/2$, typical of critical and
ultracritical particles. According to our classification, $k$ lies in region
II since (\ref{89}) gives us $\dfrac{p-q}{2}+1=1\leq k=1<\dfrac{p+1-q/2}{2}=%
\dfrac{5}{4}$. According to Table \ref{reg_tab}, this gives us the 2-nd line
(subcritical particle) with $s=c=1$ and $n_{1}=-\dfrac{1}{2}$. Thus, we see
that the force diverges.

If $\tilde{E}\neq \omega _{H}\tilde{L}+e(A_{t})_{H}(1-a\omega _{H})$, $%
\mathcal{X}_{H}\neq 0$ that gives us $s=0$. From (\ref{xi_exp}) we, as
before, deduce that $s_{1}=1$. In this case, using (\ref{n1_u}) we obtain $%
n_{1}=-\dfrac{3}{2}$, so acceleration is divergent. Thus for ultraextremal
horizons both subcritical and usual particles experience infinite forces in
the OZAMO frame.

\subsection{Verifying results}

Now let us verify our predictions explicitly. In \cite{el-geo,el-geo2}, a
reader can find equations of trajectory of particle in Kerr-Newman-(anti-)de
Sitter space-time:

\begin{equation}
\dfrac{dr}{d\tau }=-\dfrac{\sqrt{R(r)}}{r^{2}},
\end{equation}%
where for equatorial motion 
\begin{equation}
R(r)=\Big(\Big(1+\frac{\Lambda }{3}a^{2}\Big)(E(r^{2}+a^{2})-aL)+eQ\Big)%
^{2}-\Delta _{r}\Big(\Big(1+\frac{\Lambda }{3}a^{2}\Big)(aE-L)^{2}+m^{2}r^{2}%
\Big).
\end{equation}

In the non-extremal case $p=q=1$ we have $\Delta _{r}\approx \alpha v$,
where $\alpha $ is a constant. Calculating acceleration (using (\ref{a_r_o})
and \ref{a_zamo}) we find%
\begin{equation}
a^{r}=\dfrac{3eQ}{r_{h}^{4}(3a^{2}+\Lambda )^{2}}\Big((3a^{2}+\Lambda
)(E(a^{2}+r_{h}^{2})-aL)-3eQr_{h}\Big)+O(v).
\end{equation}%
We see that $a^{r}$ is finite. However, in the OZAMO frame $a_{o}^{(r)}=%
\dfrac{a^{r}}{\sqrt{A}}\sim \dfrac{1}{\sqrt{v}}$, so it diverges as we
concluded in Section \ref{non-extr}.

For the extremal case $p=q=2$ we $\ $have $\Delta \approx \alpha v^{2}$, and 
\begin{gather}
a^{r}=\dfrac{3eQ}{r_{h}^{4}(3a^{2}+\Lambda )^{2}}\Big((3a^{2}+\Lambda
)(E(a^{2}+r_{h}^{2})-aL)-3eQr_{h}\Big)+ \\
+\dfrac{3eQ}{r_{h}^{5}(3a^{2}+\Lambda )^{2}}\Big(9eQr_{h}-2(3a^{2}+\Lambda
)(E(2a^{2}+r_{h}^{2})-2aL)\Big)v+O(v^{2}).
\end{gather}

If $E=\dfrac{aL}{r^{2}+a^{2}}+\dfrac{eQr}{r^{2}+a^{2}}\dfrac{1}{1+\frac{%
\Lambda }{3}a^{2}},$ this is equivalent to the condition $E=\omega
_{H}L-eA_{H}(1-a\omega _{H})$ that gives the critical trajectory, only the
second term survives, so $a^{r}\sim v$. Thus $a_{o}^{(r)}=\dfrac{a^{r}}{%
\sqrt{A}}=O(1)$. If $E\neq \dfrac{aL}{r^{2}+a^{2}}+\dfrac{eQr}{r^{2}+a^{2}}%
\dfrac{1}{1+\frac{\Lambda }{3}a^{2}}$ that gives rise to a usual trajectory, 
$a^{r}=O(1)$. Thus we have $a_{o}^{(r)}=\dfrac{a^{r}}{\sqrt{A}}\sim \dfrac{1%
}{v}$. So, as we concluded in \ref{extr}, for critical particles
acceleration in the OZAMO frame is finite, while for usual one it diverges.

In the ultraextremal case $p=q=3$ we have $\Delta _{r}\approx \alpha v^{3} $
and we have {\normalsize 
\begin{gather}
a^{r}=\dfrac{3eQ}{r_{h}^{4}(3a^{2}+\Lambda )^{2}}\Big((3a^{2}+\Lambda
)(E(a^{2}+r_{h}^{2})-aL)-3eQr_{h}\Big)+ \\
+\dfrac{3eQ}{r_{h}^{5}(3a^{2}+\Lambda )^{2}}\Big(9eQr_{h}-2(3a^{2}+\Lambda
)(E(2a^{2}+r_{h}^{2})-2aL)\Big)v+O(v^{2}).
\end{gather}%
}

We see that if $E=\dfrac{aL}{r^{2}+a^{2}}+\dfrac{eQr}{r^{2}+a^{2}}\dfrac{1}{%
1+\frac{\Lambda }{3}a^{2}}$ that is equivalent to the condition $E=\omega
_{H}L-eA_{H}(1-a\omega _{H})$. As now $s=c=1$\thinspace \thinspace $<\frac{p%
}{2}$, this and corresponds to the subcritical trajectory (line 2 in Table %
\ref{class_of_tr}). In the above formula for $a^{r}$ only the second term
survives, so $a^{r}\sim v$. Thus $a_{o}^{(r)}=\dfrac{a^{r}}{\sqrt{A}}\sim 
\dfrac{1}{\sqrt{v}}$. If $E\neq \dfrac{aL}{r^{2}+a^{2}}+\dfrac{eQr}{%
r^{2}+a^{2}}\dfrac{1}{1+\frac{\Lambda }{3}a^{2}}$, $a^{r}=O(1)$ and $%
a_{o}^{(r)}=\dfrac{a^{r}}{\sqrt{A}}\sim \dfrac{1}{v^{3/2}}$. So, for
subcritical and usual particles the acceleration in the OZAMO frame diverges.

We see that all the results for the metric under discussion agree completely
with our general scheme.

\section{Restrictions and reservations}

In the present work we considered collisions in the test particles
approximation. This means that we neglect backreaction of particles on the
metric. This is just the same approximation that was made in pioneering
works \cite{ban} - \cite{pir3}. Clearly, account of self-gravitation can
change the results qualitatively. For example, in the paper \cite{sh}
collisions of massive spherically symmetric charged shells were considered.
It was shown that the self-gravitation bounds the energy in the center of
mass frame that otherwise would be as large as one likes \cite{jl}. In doing
so, these authors found that according to their eq. (42), the factor $\eta
=\left( \frac{M_{1}}{\mu }\right) ^{1/4}\,$appears that restricts $E_{c.m.}$%
, where $M_{1}$ is the mass of a central black hole, $\mu $ being the proper
mass of the shell, $M_{1}\gg \mu $. Although the restriction does indeed
take place, this factor, being finite, is nonetheless is very large, $\eta
\gg 1$, so qualitatively the effect remains.

In \cite{hod} another approach was used for collisions of particles (not
shells) that was based on the\ hoop conjecture \cite{hoop} . Remarkably, the
same factor $\eta $ was obtained there. Quite recently, Ref. \cite{real}
appeared in which the hoop conjecture was applied to scenarios in the
background of rotating black holes, with the conclusion that this conjecture
regularizes $E_{c.m.}$ making it finite but this quantity remains quite
high. For example, these authors found that $E_{c.m.}$ can be only three
orders of magnitude less than the Planck energy.

In our view, all this is quite natural: if there exists an effect that leads
to indefinitely large values of energy, something should exist that bounds
it from above, leaving it finite but large. In this respect, the test
particle approximation used in our work (as well as almost all works on the
BSW effect) is the only first step. {}The scenarios considered in our paper
are much more involved that those in \cite{ban}, \cite{jl}, since they take
into account forces of different nature and more types of trajectories.
Therefore, in our context the effect of self-gravity is less obvious in
advance. It looked reasonable to study collisions firstly in the text
particle approximation and only afterwards to include into consideration
self-gravitation in addition to aforementioned factors. We hope to return to
this issue in our future work.

\section{Summary and conclusions\label{sum}}

Thus we constructed classification of near-horizon trajectories according to
their four-velocities and four-accelerations. We singled out so-called usual
particles (without fine-tuned parameters) and fine-tuned ones. As the
necessary condition of the BSW effect implies the process with participation
of a fine-tuned particle, the main emphasis was made on investigation of
properties of such particles. In turn, the set of fine-tuned particles is
split to subcritical, critical and ultracritical ones depending on their
near-horizon behavior. We found the conditions when the components of
acceleration remains finite for each type of a trajectory. For fine-tuned
particles, the relevant frame for measuring these components is the OZAMO
frame since a corresponding observer does not cross the horizon, similarly
to a fine-tuned particle.

The properties of the metric are characterized by the set of three numbers $%
p $, $q$, $k$ responsible for the near-horizon behavior. We also introduced
the numbers $n_{0}$, $n_{1}$, $n_{2}$ which show the rate with which the
tetrad components of the acceleration in the OZAMO frame change near the
horizon. Then, the requirement of finiteness of acceleration for fine-tuned
particles reduces to the conditions $n_{i}\geq 0$ for $i=0,1,2$. These
conditions lead to constraints in the space of relevant parameters
describing the metric. The results of our work are presented in the Tables %
\ref{class_of_tr} - \ref{tab_kregs}.

A separate interesting issue that revealed itself in the course of our
investigation is the principle of kinematic censorship. By itself, it looks
very simple or even trivial since it is obvious that in any act of collision
the energy $E_{c.m.}$ cannot be infinite. Meanwhile, as we saw it, the proof
of the fact that this principe is indeed realized in all scenarios under
study turned out quite nontrivial in our context. We showed that kinematic
censorship is indeed preserved. Namely, either (i) the proper time required
to reach the horizon for a fine-tuned particle participating in the BSW
process is infinite or (ii) the force diverges, or (iii) the horizon fails
to be regular. Actually, this principle is a power tool that enables one to
select between possible and forbidden scenarios, even without having
explicit solutions of particle motion.

To verify the obtained results, we checked them using the
Kerr-Newman-(anti-)de Sitter metric as an example.

Although our main motivation was connected with the study of the BSW effect,
the obtained results for the relationship \ between the type of trajectory
and acceleration can be of some use in more general contexts.

It is of interest to extend the present results to nonequatorial motion, the
BSW processes with circle orbits and near-critical particles. Also, it would
be interesting to take into account the effects of self-gravitation briefly
mentioned in the preceding section,

\appendix

\end{document}